\documentclass[graphics,floatfix, footinbib,tightenlines,nobibnotes, aps, pra, onecolumn,notitlepage,superscriptaddress]{revtex4-1}
\usepackage{amsmath,amssymb,wasysym}
\usepackage{graphicx}
\usepackage{dcolumn}
\usepackage{bm}
\usepackage{braket}
\usepackage{subcaption}
\usepackage{verbatim}
\usepackage{float}
\usepackage{bbold}
\usepackage{listings}
\usepackage{color}
\usepackage{xcolor}
\usepackage{relsize}
\usepackage{amsthm}
\usepackage{enumerate}
\usepackage{accents}
\usepackage{dsfont}
\usepackage{soul,xcolor}
\usepackage[T1]{fontenc}
\usepackage[colorlinks=true ,urlcolor=blue,urlbordercolor={0 1 1}]{hyperref}

\newcommand{\beq}{\begin{eqnarray}}
\newcommand{\eeq}{\end{eqnarray}}

\usepackage{mathtools}

\newcommand{\bsp}{\begin{split}}
\newcommand{\esp}{\end{split}}

\graphicspath{{./images/}}
\def\bea{\begin{eqnarray}}
\def\eea{\end{eqnarray}}

\makeatletter

\def\env@sqcases{%
  \let\@ifnextchar\new@ifnextchar
  \left\lbrack
  \def\arraystretch{1.2}%
  \array{@{}l@{\quad}l@{}}%
}
\makeatother
\definecolor{blue(ncs)}{rgb}{0.0, 0.53, 0.74}

\begin{document}

\setstcolor{red}

\title{Theory of Sondheimer magneto-oscillations beyond semiclassical limit}
\author{Alexander Nikolaenko}
\affiliation{Department of Physics, Harvard University, Cambridge, MA 02138, USA}
\author{Pavel A. Nosov}
\affiliation{Department of Physics, Harvard University, Cambridge, MA 02138, USA}

\date{\today}

\begin{abstract}
In conducting films subjected to an out-of-plane magnetic field, electron motion along the field direction gives rise to conductance oscillations periodic in field intensity - a phenomenon known as Sondheimer oscillations. Traditionally, these oscillations were understood within the semiclassical framework of kinetic theory. However, their behavior in the quantum regime (i.e. at strong fields and weak disorder) remains unclear, particularly due to potential interference with quantum Shubnikov-de Haas magneto-oscillations. In this work, we develop a comprehensive theory of quantum magnetoconductivity oscillations in metallic films of finite thickness, fully capturing the interplay between the Sondheimer and Shubnikov–de Haas effects beyond the semiclassical limit. By treating surface scattering, in-plane Landau quantization, and dimensional confinement along the magnetic field direction on equal footing, we reveal an intricate hierarchy of oscillation patterns and characterize how their amplitudes and frequencies depend on various physical parameters. Our results pave the way for systematic characterization of thin metallic films with boundary-dominated transport properties.
\end{abstract}

\pacs{Valid PACS appear here}
\maketitle{}

\section{Introduction}
\noindent

Quantum effects in transport play an important role in understanding the fundamental theory of metals and in the advancement of modern technology. The physics becomes especially rich upon applying a magnetic field, since the semiclassical trajectories of electrons transform into cyclotron orbits and the band structure reconstructs into Landau levels \cite{Shoenberg_1984,abrikosov2017fundamentals}. Furthermore, when conducting films are thin, the geometric effects of the boundaries become relevant \cite{Ashcroft1988,BRANDLI196961}.
In recent years, significant progress has been made in fabricating thin samples with a long mean-free path and generating strong magnetic fields \cite{putzke2020h}. These conditions are ideal for observing both geometrical and quantum effects.

A pioneering semiclassical treatment of transport in thin films was developed by Fuchs \cite{Fuchs1938}, who showed that a film’s conductivity decreases once its thickness $d$ becomes comparable to the mean free path $l$.
Fuchs' theory was later generalized by Sondheimer \cite{Sondheimer1952} to include the effects of a magnetic field. Sondheimer showed that in a perpendicular magnetic field with diffusive surface scattering, the conductivity exhibits oscillations that vary {\it{linearly}} with the magnetic field $B$. These oscillations, referenced later as Sondheimer oscillations, are a semiclassical geometric effect associated with how many cyclotron orbits the electron completes between the consecutive collisions with the surfaces. Interest in such boundary-induced magneto-oscillations has recently revived \cite{Taen2023, Kim2011,Delft2021, Mallik2022, PhysRevB.108.245405}, especially in systems where conventional Shubnikov–de Haas (SdH) oscillations are weak or inaccessible.

Although Sondheimer's theory accurately predicted the emergence of linear-in-$B$ oscillations in many materials \cite{Grenier1966}, its limitations are clearly evident. First, it breaks down in the regime of very thin films $d \ll l$, where quantization effects along the direction of the field become significant. In fact, in the limit $l \rightarrow \infty$, the conductivity exhibits an unphysical logarithmic divergence at small fields. Second, it is unable to describe the regime of the strong magnetic field $\omega_c \tau \gtrsim1$ (here $\omega_c$ is the cyclotron frequency, and $\tau$ is the mean free time) such that individual Landau levels become discernible, leading to the emergence of SdH magneto-oscillations.
Furthermore, some claims about the amplitude and phase of the oscillations \cite{Grenier1966} implicitly assume the strong field limit, $\omega_c \tau \gg 1$, rendering the semiclassical theory not entirely consistent.

Over the years, many authors have attempted to extend Fuchs's and Sondheimer's theory to the quantum limit of ultra-thin samples \cite{Munoz2017}. One approach that successfully unified both semiclassical and quantum limits in the absence of a magnetic field was proposed by Sheng et al. \cite{Sheng1995}. The authors considered an electron gas confined between two potential barriers and calculated Green's functions in the presence of bulk impurities and rough surface scattering. They further used Kubo's linear response formalism to compute conductivity in the ultrathin limit. Their answer no longer exhibited an unphysical divergence in the limit $l \rightarrow \infty$ and reduced to the semiclassical formula when the discrete nature of the electron states could be safely neglected, assuming angle-independent reflectivity. Surface scattering effects in thin metallic films remain an active area of research \cite{Chatterjee2010,KETENOGLU20123828,Lingjun2015,Tianji2018,Nair2021}, but most developments to date have focused on the $B=0$ limit.

In this work, we expand on previous studies by considering an electron gas in a perpendicular magnetic field and extending the theory beyond Sondheimer's semiclassical analysis. We model surface roughness via a random potential at the boundary with an effective scattering length $\delta$, and derive transport properties from the Kubo formula treated within the self‑consistent Born approximation. In the case of many occupied Landau levels we obtain a closed expression for the oscillatory conductivity that reduces to Sondheimer's formula in the limit of thick samples $d \gg l$ and small magnetic fields $\omega_c \tau \ll 1$.  Our theory also eliminates an unphysical divergence of conductivity in the ultra-thin boundary-dominated regime $l \gg d$, coincides with Sheng et al.'s formulas in the zero field limit, and demonstrates negative quadratic magnetoconductivity at small fields. In the limit of strong magnetic fields ($\omega_c \tau \gg 1$), additional quantum oscillatory contributions emerge, which could be associated with a version of quantum SdH oscillations, modified by the quantization of modes along the magnetic field direction. The latter effect results in a series of additional harmonics with $B$‑linear phase shift of magnitude $\propto d^{2}/ l_{B}^{2}$, where $l_B$ is the magnetic length. In addition to the exponential Dingle factor originating from bulk disorder, these harmonics are further suppressed when boundary scattering is diffusive. We also investigate the limit of a few filled Landau levels, where the $k_z$-momentum quantization becomes the dominant source of oscillations which are neither strictly $B$ nor $1/B$ periodic.

In addition, we study the temperature dependence of both types of oscillations. Due to their semiclassical nature, Sondheimer oscillations are exponentially suppressed only when the number of oscillations exceeds $\epsilon_F/T$, where $\epsilon_F$ is the Fermi energy. Since in most metallic films $\epsilon_F \gg T$, these oscillations are remarkably robust against thermal smearing. By contrast, the quantum oscillatory contributions are exponentially suppressed once $T \gg \omega_c$. Finally, the geometric contributions with linear-in-$B$ phase shifts decay at temperatures $T\gg \epsilon_F/(k_F d)$ exceeds the individual level spacing associated with the quantized $k_z$ momentum.

We note that, beyond their distinct physical origins, Sondheimer and quantum oscillations probe different aspects of the Fermi surface. SdH oscillations provide information about the extremal cross-sectional areas of the Fermi surface perpendicular to the applied field, whereas Sondheimer oscillations are sensitive to the orbits with maximum $k_z$ momentum, thereby probing the poles of the Fermi surface. Observing both not only bridges the semiclassical and quantum regimes but also provides a powerful tool for fermiology, enabling access to complementary parameters typically inaccessible in a single measurement.

Experimentally, Sondheimer oscillations were observed in various single crystals shortly after Sondheimer's original prediction \cite{Babiskin1957,Grenier1966,phonon}. Most of these early measurements were performed in the dirty limit, where quantum SdH oscillations could not be resolved. Therefore, semiclassical theory within the Boltzmann formalism was sufficient to explain most of the experimental findings. More recently, advances in fabricating clean ultrathin samples of graphene and semimetals have enabled the simultaneous observation of Sondheimer oscillations and quantum SdH oscillations \cite{Taen2023, Kim2011,Delft2021, Mallik2022, PhysRevB.108.245405}. 
Thus, the natural question arises how to incorporate both types of oscillations under a unified theory. The formalism developed in this work provides a theoretical framework for understanding the transport regime in which both quantum and semiclassical oscillations coexist and become simultaneously relevant.

The paper is structured as follows: in Section~\ref{sec:sondh} we revisit the semiclassical derivation of Sondheimer oscillations using the Boltzmann formalism and extend it to finite temperatures. We also discuss the limitations of the semiclassical theory and the conditions under which it breaks down. In Section~\ref{sec:model}, we formulate the model, examine the relevant parameter regime, and introduce the Green's function formalism. In Section~\ref{sec:self_energy}, we derive Dyson's equation and the Born approximation for the self-energy and solve them within the specified limit. We use Kubo's linear response formalism to compute the conductivity in Section~\ref{sec:cond} and analyze various limiting cases. We thoroughly discuss the main formulas in Section~\ref{sec:disc} and provide the physical intuition behind the results. This section is also intended for readers who wish to avoid the heavy formalism of the previous sections and instead seek a concise summary of the work.
Finally, Section~\ref{sec:concl} concludes by connecting our results to recent experiments. We leave some technical details to Appendices \ref{app:spat_osc}, \ref{sec:appendix_few_LLs}, while additional details regarding the case with non-parabolic dispersion and its relation to experiments are discussed in Appendix \ref{app:sondh}.

\section{Semiclassical Sondheimer theory revisited}
\label{sec:sondh}
\begin{figure}[h]
    \center{\includegraphics[width=0.45\linewidth]{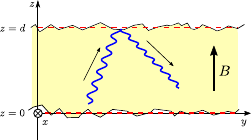}}  
\caption{Schematic representation of a metallic thin film of thickness $d$ with rough surfaces and in a perpendicular magnetic field. The red dashed lines denote the average position of the boundary, and the blue solid line represents a classical electron trajectory involving both cyclotron motion and surface scattering.
}
\label{fig:setup}
\end{figure}

We start this section by reviewing Sondheimer's semiclassical derivation of the electrical conductivity. Consider an electron gas confined between two rough surfaces located at $z=0$ and $z=d$, subject to an electric field $\vec{E}=(E_x,E_y,0)$ and a perpendicular magnetic field $\vec{B}=(0,0,B)$, see Fig.~\ref{fig:setup}.
The quasi-free electrons with the dispersion relation $\epsilon_k=k^2/(2m)-\mu$, where $m$ is the mass and $\mu$ is the chemical potential, are described by the distribution function $f(\vec{k},\vec{r})$. Due to the system's translational invariance in the $x$ and $y$ directions, and assuming that the surface roughness is, on average, captured by the boundary conditions, the distribution function depends only on the $z$-coordinate.
The Boltzmann transport equation in the relaxation time approximation is given by
\begin{equation}
\left( \vec{E}+\vec{v}\times \vec{B} \right)\frac{\partial f}{\partial \vec{k}}+\vec{v} \frac{\partial f}{\partial \vec{r}}=-\frac{f-f_0}{\tau},    
\end{equation}
where $\tau$ denotes the relaxation time and $f_0(v)=1/(1+\exp((m v^2/2-\mu)/T))$ represents the equilibrium Fermi-Dirac distribution function. We also adopt units where $c{=}\hbar{=}k_B{=}e^2{=}1$. For a quadratic dispersion relation, the velocity and momentum are simply proportional, given by $\vec{v}=\vec{k}/m$. Assuming the electric field is weak, we linearize the distribution function around its equilibrium value $f(\vec{v},z)=f_0(v)+(c_x(v,v_z,z) v_x+c_y(v,v_z,z) v_y)\partial f_0/\partial v$, where $v=|\vec{v}|$ and terms of order $O(E^2)$ are neglected.
Setting the terms proportional to $v_x$ and $v_y$ to zero independently, we obtain
\begin{equation}
    \begin{dcases}
        &\frac{E_x}{m v}-\frac{B}{m}c_y+v_z \frac{\partial c_x}{\partial z}=-\frac{c_x}{\tau}\\
          &\frac{E_y}{m v}+\frac{B}{m}c_x+v_z \frac{\partial c_y}{\partial z}=-\frac{c_y}{\tau}.\\
    \end{dcases}
    \label{eq:sondh_cx_cy}
\end{equation}
The two equations can be combined into a single one by complexifying the variables. Defining $E=E_x + i E_y$ and $c=c_x+i c_y$ and solving the first-order linear differential equation, we have
\begin{equation}
   c(v,v_z,z)=-\frac{E \tau}{m v}\frac{1}{1+i B \tau/m} \left( 1+A(v,v_z) e^{-\frac{z}{\tau v_z}(1+i B \tau/m)}\right).
\end{equation}
The proportionality constant $A(v,v_z)$ is determined from the boundary conditions. Fully diffusive scattering is modeled by imposing the condition $f(v_z<0,z=d)=f(v_z>0,z=0)=f_0$ for electrons reflecting from the boundaries. Specular scattering, on the other hand, is modeled by requiring that the distribution function of incoming electrons equals that of outgoing electrons at the boundary: $f(v_z,z=0,d)=f(-v_z,z=0,d)$. In the most general case, a fraction $0<R<1$ of electrons is scattered specularly, while the rest is scattered diffusively and the boundary conditions take the form
\begin{equation}
    \begin{dcases}
        &f(v_z,z=0)=R f(-v_z,z=0)+(1-R)f_0, v_z>0\\
          &f(v_z,z=d)=R f(-v_z,z=d)+(1-R)f_0, v_z<0.\\
    \end{dcases}
\end{equation}

 The local current in the $x$-direction is expressed as
\begin{equation}
    J_x(z)=\frac{m^3}{(2\pi)^3} \int dv_x dv_y dv_z f(\vec{v},z) v_x=\frac{m^3}{(2\pi)^3} \int dv_x dv_y dv_z c_x(v,v_z,z) v_x^2 \frac{\partial f_0 }{\partial v}.
\end{equation}
At zero temperature ($T=0$) the Fermi-Dirac distribution function reduces to a step function with its derivative given by $\partial f_0/\partial v=-\delta(v-v_F)$. This implies that only states close to the Fermi surface contribute to transport. 

We convert the integral to spherical coordinates and integrate over the azimuthal angle. We further compute total current by integrating the local one over $z$-direction.
Finally, by making the change of variables, we can establish the connection to Sondheimer's formula
\begin{equation}
    \sigma^{xx}=\frac{d}{l}\sigma_0 \Re\left( \frac{1}{\zeta}-\frac{3}{2\zeta^2}(1-R)\int_1^{\infty} dt \left( \frac{1}{t^3}-\frac{1}{t^5}\right)\frac{ 1-e^{-\zeta t}}{1-Re^{-\zeta t}}\right),
    \label{eq:sondh_full}
\end{equation}
 where $\sigma_0=n \tau/m$ is the three-dimensional conductivity in the absence of a magnetic field with a total number of particles $n= m^3 v_F^3/(6 \pi^2)$,$l=v_F \tau$ is the mean free path, $\zeta=\frac{d}{l}(1+i \omega_c \tau)$ is a dimensionless variable which we use frequently throughout the paper,  $\omega_c=B/m$ is the cyclotron frequency, and $\Re$ denotes the real part.

Next, we proceed with an analysis of the formula in various regimes. First, we consider the case with no magnetic field ($B=0$) and examine different film thicknesses: 
\begin{equation}
    \sigma^{xx}=
    \begin{dcases}
          & \sigma_0 \left( 1-\frac{3}{8}(1-R)\frac{l}{ d}\right), \quad d \gg l\\
            &   -\frac{3}{4}\sigma_0\frac{1+R}{1-R} \frac{d}{l} \log \frac{d}{l} , \quad \; \; \; d \ll l.\\
    \end{dcases}
\end{equation}

In the limit of large, but finite thickness, the conductivity decreases slightly from its bulk value $\sigma_0$. This reduction is stronger when the surface scattering is diffusive. This effect was originally predicted by Fuchs \cite{Fuchs1938}. In the opposite limit of ultrathin samples, the conductivity goes to zero as $\sigma_{xx} \propto d \log d$. 

An unphysical result arises when taking the limit of ultraclean samples with the mean free path $l \rightarrow \infty$. In this case, $\sigma_0 \propto l$ and $\sigma^{xx}$ diverges logarithmically, $\sigma^{xx}\propto \log l$. This behavior originates from from electrons moving at arbitrarily small angles relative to the surface. In the absence of bulk impurities, such electrons require an infinite time to relax their momentum at the boundaries, leading to a divergence of the conductivity. This limitation of semiclassical theory was pointed out by Sheng et al. \cite{Sheng1995}. The discrepancy disappears in a quantum treatment, since the $k_z$ momentum becomes quantized. Starting in Section \ref{sec:model}, we develop a quantum formalism that properly generalizes the semiclassical description for thin, high-purity samples.

In the case of a finite magnetic field, the non-oscillatory part of the conductivity is given by
\begin{equation}
    \sigma^{xx}=
    \begin{dcases}
          & \frac{\sigma_0}{1+\omega_c^2 \tau^2} \left( 1-\frac{3}{8}(1-R)\frac{l}{ d} \frac{1-\omega_c^2 \tau^2}{1+\omega_c^2 \tau^2}\right), \quad d \gg l\\
            &  -\frac{3}{4}\sigma_0 \frac{d}{l}\frac{1+R}{1-R} \log\left(\frac{d}{l}\sqrt{1+\omega_c^2 \tau^2}\right),\quad  \quad \; \frac{d}{l}\sqrt{1+\omega_c^2 \tau^2} \ll 1-R.\\
    \end{dcases}
    \label{eq:sigma_class_finiteB}
\end{equation}
When $l \rightarrow\infty$ we once again encounter an unphysical result: as the magnetic field $B$ decreases, the conductivity diverges logarithmically. We could also take into account the oscillatory effects of exponential terms in Eq.~(\ref{eq:sondh_full}).  Performing the integration by parts twice and assuming $d\gg l$, $|\zeta| \gg 1$, we obtain additional oscillatory corrections to the conductivity:

\begin{equation}
 \sigma^{xx}_{\rm osc}=\sigma_0\frac{3 (l/d)^3}{(1+\omega_c^2 \tau^2)^2}e^{-d/l} (1-R)^2\cos \left(\omega_c \tau\frac{ d}{l} +4 \arctan (\omega_c \tau)\right)  .
    \label{eqn:osc}
\end{equation}
These oscillations were first introduced by Sondheimer \cite{Sondheimer1952} and later observed in various materials \cite{Grenier1966,Alstadheim1968,Delft2021,Taen2023}. In contrast to the non-oscillatory part in Eq.~(\ref{eq:sigma_class_finiteB}), which decays as $B^{-2}$, the amplitude of oscillations decays more rapidly as $B^{-4}$ with increasing magnetic field. Note that the power of the decay is not universal and could depend on the topology of the Fermi surface, see \cite{gurevich1959} and Appendix \ref{app:sondh} for further discussions. We also emphasize that diffusive boundary scattering is essential for the existence of Sondheimer oscillations, as Eq.~\eqref{eqn:osc} vanishes in the specular limit $R=1$. Most importantly, the frequency of oscillations in Eq.~\eqref{eqn:osc} is $B d/k_F$, where $k_F=m v_F$ is the Fermi momentum. It does not depend on the bulk disorder and is linear in $B$ as opposed to the $1/B$ dependence in the Shubnikov-de Haas effect which arises from Onsager quantization \cite{Shoenberg_1984}. Therefore, we stress once more that Sondheimer oscillations are not a quantum effect; rather, they arise purely from the finite thickness of the film and the diffusive nature of electron reflection at the surfaces.

We now turn to the applicability of Sondheimer's result. First, as noted previously, we should be in the thick film limit ($d \gg l)$. Second, the magnetic field should be weak enough that individual Landau levels remain unresolved ($\omega_c \tau \ll 1$). These stringent conditions make experimental detection of the oscillations challenging, as the signal is suppressed by the factor $e^{-d/l}$. In practice, experimentalists often relax one of the conditions, allowing either a strong magnetic field\cite{Grenier1966} or thin samples with $d \sim l$ \cite{guo2025}. Therefore, developing a complete quantum theory of Sondheimer oscillations is important from both theoretical and experimental perspectives. We proceed with formulating such a quantum model in the next section.

Another limit worth considering is $\tau \rightarrow \infty$. In this limit $\zeta= i d \omega_c/v_F$ does not depend on $\tau$ so Eq.~(\ref{eq:sondh_full}) gives a finite $\sigma^{xx}$:
\begin{equation}
    \sigma^{xx}=  \frac{3 n v_F}{2d m \omega_c^2}(1-R) \Re\int_1^{\infty} dt \left( \frac{1}{t^3}-\frac{1}{t^5}\right)\frac{ 1-e^{-i d \omega_c t/v_F}}{1-Re^{-id \omega_c t/v_F}}.\label{eq:Sigma_xx_1/tau=0_classical}
\end{equation}
The conductivity becomes zero once the boundary scattering is fully specular. Note that when $|\zeta| \gg 1$, which is equivalent to having many oscillations, Eq.~(\ref{eqn:osc}) still holds.

\begin{figure}[t!]
      \begin{minipage}[h]{1\linewidth}
    \center{\includegraphics[width=0.45\linewidth]{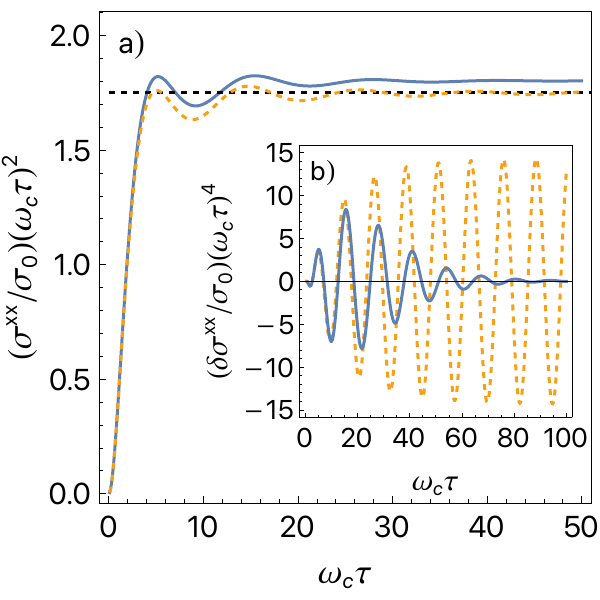}} 
    \end{minipage} 
\caption{(a) Longitudinal conductivity at zero temperature (orange dashed line) and at finite temperature (blue line). The parameters used are: $d/l=0.5$, $T/\epsilon_F=0.12$, $R=0$.
(b) Inset shows only oscillatory part of the conductivity, multiplied by $(\omega_c \tau)^4$. The oscillations at finite temperature decay at large magnetic fields.}
\label{fig:Sondheimer_classical}
\end{figure}

The semiclassical treatment outlined above can be readily extended to finite temperatures. In this case, Eq.~(\ref{eq:sondh_full}) includes an additional integral over dimensionless velocity $u$:
\begin{equation}
    \sigma^{xx}= \frac{d}{l}\sigma_0\Re \int du \frac{\partial f_0}{\partial u}u^2 \left( \frac{u}{\zeta}-\frac{3u^2}{2\zeta^2}(1-R)\int_1^{\infty} dt \left( \frac{1}{t^3}-\frac{1}{t^5}\right)\frac{ 1-e^{-\zeta t/u}}{1-Re^{-\zeta t/u}}\right),
    \label{eq:sondh_full_T}
\end{equation}
where $v=v_F u$. Several typical examples of the resulting oscillations are depicted in Fig.~\ref{fig:Sondheimer_classical}(a). The oscillating part of $\sigma^{xx}$ can be evaluated analytically assuming both $|\zeta| \gg 1$ and $T/\epsilon_F\ll 1$. After defining $u=1+\xi$  and expanding around $\xi\approx 0$ we obtain:

\begin{equation}
      \sigma^{xx}_{\rm osc}=3\sigma_0  (l/d)^3 (1-R)^2\;\Re\;\frac{e^{-\frac{d}{l}(1+i \omega_c \tau)}}{(1+i \omega_c \tau)^4} \frac{\frac{\pi d T}{2l\epsilon_F }(1+i \omega_c \tau)}{\sin\left(\frac{\pi d T}{2l\epsilon_F }(1+i \omega_c \tau)\right)}\;.
\end{equation}

The amplitude of oscillations at finite temperature becomes non monotonic, see Fig.~\ref{fig:Sondheimer_classical}(b). In the limit $\omega_c \tau {\gg} 1$ the last factor resembles the Lifshitz-Kosevich factor $\propto X/\sinh(\pi X/2)$, but with the distinct argument $X= (d \omega_c/v_F)(T/\epsilon_F)$. The Sondheimer oscillations are therefore exponentially suppressed when $d \omega_c/v_F \gg \epsilon_F/T$. Note, that thermal suppression happens at large magnetic fields, rather than at small fields, as in the ordinary Shubnikov–de Haas oscillations. This is because the magnetic field enters Lifshitz-Kosevich factor linearly, rather than through $B^{-1}$ in the SdH case. In particular, at strong magnetic fields and finite temperature the amplitude acquires an additional exponential factor $B^{-3} e^{-B/B_0}$, where $B_0=2 m v_F \epsilon_F/(\pi  d T)$.

The physical intuition behind this unusual magnetic field dependence at finite temperature is as follows. Sondheimer oscillations reflect how many cyclotron revolutions an electron makes before being scattered by the sample boundaries. The temperature naturally broadens the velocity distribution and the contribution from electrons completing $n$ cyclotron orbits is out of resonance with contribution from electrons completing $n+1$ cyclotron orbits. Once the magnetic field becomes strong, an electron takes less time to complete one cyclotron orbit. Therefore, even small broadening of the velocities results in strong mixing between different trajectories and supresses the oscillations.

In any metallic film $\epsilon_F/T \gg 1$, and so we expect many oscillations before observing any decay. This makes Sondheimer oscillations more resistant to temperature than quantum (Shubnikov–de Haas) oscillations, which are typically suppressed for $T\gg \omega_c$ \cite{Shoenberg_1984,abrikosov2017fundamentals}.
With applications to realistic materials in mind, in Appendix \ref{app:sondh} we also present several additional generalizations of the semiclassical Sondheimer theory, including anisotropic scattering rates $\tau_\parallel$ and $\tau_\perp$ and more complex Fermi surfaces (the latter extension was originally discussed by Gurevich in \cite{gurevich1959}).

In summary, the quasi-classical theory developed by Fuchs and Sondheimer provides a good description in the "double semiclassical" limit of thick films ($d\gg l$) and weak magnetic fields ($\omega_c \tau \ll1$). However, once the disorder is weak, the samples are thin, or the magnetic field is strong, the quasiclassical theory often yields unphysical results.  In these regimes, quantum effects become significant, and a full quantum treatment is needed to describe the transport properties. The corresponding theory will be developed in the following sections.

\section{Model and relevant scales}
\label{sec:model}
We consider a normal metal thin film with rough surfaces, in the presence of bulk impurities and a perpendicular magnetic field, see Fig.~\ref{fig:setup}. The film is confined in the $z$ direction and extends in the $(x,y)$ plane (with $r_\parallel$ denoting in-plane coordinate vectors). The boundaries of the film (with positions $z_{0,1}(r_\parallel)$) are random such that $\langle z_\sigma(r_\parallel) \rangle =\sigma d$, $\sigma=0,1$, and the variance is assumed to be much smaller than $d$. The thickness $|z_0(r_\parallel) -z_1(r_\parallel)|$ equals to $d$ on average. The resulting single-particle Hamiltonian reads as
\begin{equation}\label{eq:H_full_1}
    \mathcal{H} = \frac{1}{2m}\left(-i\nabla-e\bm{A}\right)^2 +v(r) + V 
    \Big[\theta(z-z_{1}(r_\parallel))+\theta(z_{0}(r_\parallel)-z)\Big]\;.
\end{equation}
Here $m$ is the electron mass, $\bm{A}$ is the vector potential corresponding to the static perpendicular magnetic field $B$ (we use the Landau gauge: $A_y=-Bx$, $A_x=A_z=0$), and $v(r)$ is a random Gaussian potential representing weak impurities within the film's bulk, such that
\begin{equation}\label{eq:v_average}
    \langle v(r)\rangle =0\;,\quad   \langle v(r)v(r')\rangle =v_0^2 \delta(r-r')\;,
\end{equation}
where the angle brackets stand for disorder averages. The last term in Eq.~\eqref{eq:H_full_1} is a boundary confining potential which vanishes within the bulk defined by the given realization of surface positions, and is equal to $V$ outside the film (the limit $V\rightarrow \infty$ is assumed). The most systematic way of handling surface roughness is to perform a unitary transformation using the dilation operator in order to "straighten out" the boundary at the price of generating additional terms stemming from the non-commutativity of the kinetic energy and local dilations \cite{Maekawa1986,Ashcroft1988}. However, when the boundaries fluctuate around their average positions only slightly, the net effect of this transformation can be understood by simply expanding the confining potential with $z_{\sigma}(r_\parallel)=\sigma d+\eta_\sigma(r_\parallel)$  (here $\sigma=0,1$) at the lowest-order in $\eta_{\sigma}(r_\parallel)$ \cite{Fishman,Fishman2}. The resulting modified Hamiltonian acquires a simple form
\begin{equation}\label{eq:H_full_2}
  \mathcal{H}' =\frac{1}{2m}\left(-i\nabla-e\bm{A}\right)^2 + V 
    \Big[\theta(z-d)+\theta(-z)\Big]+ v(r) + V\sum\limits_{\sigma=0,1}\eta_\sigma(r_\parallel) \delta(z-\sigma d)\;.
\end{equation}
Here $\eta_\sigma(r_\parallel)$ can be now viewed as the effective roughness potential on each of the boundaries. For simplicity, we assume that it is distributed as a  "white-noise" Gaussian potential with the correlation functions given by
\begin{equation}\label{eq:eta_average}
    \langle  \eta_\sigma(r_\parallel)\rangle=0\;,\quad \langle\eta_\sigma(r_\parallel)\eta_{\sigma'}(r_\parallel')\rangle = \delta_{\sigma\sigma'}\eta_0^2 \delta(r_\parallel-r'_\parallel)\;.
\end{equation}
As will become clear shortly, correctly capturing boundary scattering requires keeping the potential $V$ in both boundary terms of Eq.~\eqref{eq:H_full_2} arbitrarily large but finite throughout certain intermediate steps of the derivation. The limit $V \rightarrow \infty$ should only be taken at the final stage of the calculation.

Our model involves multiple competing length and energy scales, making it useful to define several dimensionless parameters — some of which already appeared in the preceding section —  that characterize distinct physical regimes. The first is $d/l$, where $l$ is the bulk mean free path. This parameter governs the importance of boundary-induced quantization effects. In the limit $d/l \gg 1$, electron motion along the $z$-direction becomes semiclassical and diffusive, leading to effectively quasi-3D behavior in observable quantities. The second key parameter is $\omega_c \tau$, where $\omega_c = B/m$ is the cyclotron frequency and $\tau = l/v_F$ is the impurity-induced relaxation time, with $v_F$ denoting the Fermi velocity. (We adopt units where $\hbar = c = k_B = 1$.) This dimensionless ratio controls the extent to which Landau quantization is smeared by disorder. When $\omega_c \tau \ll 1$, conventional Shubnikov--de Haas (SdH) oscillations are exponentially suppressed, and only semiclassical in-plane dynamics remain. Thus, the leading-order behavior in a double expansion in $l/d$ and $\omega_c \tau$ corresponds to the ``double semiclassical'' limit captured by Sondheimer’s theory (see Sec.~\ref{sec:sondh}). A third important parameter is $\omega_c / \epsilon_F$, where $\epsilon_F = m v_F^2 / 2$ is the Fermi energy. This ratio determines the number of occupied Landau levels. We first assume $\omega_c / \epsilon_F \ll 1$, ensuring that many Landau levels are filled and after that consider an opposite limit of a few Landau levels with $\omega_c / \epsilon_F \sim 1$. Rough surface scattering is characterized by the dimensionless parameter $k_F \delta \sim k_F^4 \eta_0^2$, where $k_F$ is the Fermi momentum (a more precise definition of the length scale $\delta$ given below). This parameter quantifies the degree of momentum relaxation: $k_F \delta = 0$ corresponds to perfectly specular scattering, while $k_F \delta \sim 1$ indicates fully diffusive scattering. Finally, the dimensionless parameter $k_F d$ determines the number of modes in $z$-direction. Although our main focus is on the regime where $k_F d
\gg 1$ but $d/l$ is not necessarily large, we also derive results for the opposite limit of very few modes, where discretization effects become significant.

In our analysis, we will be using the simplified Hamiltonian $\mathcal{H}'$ in Eq.~\eqref{eq:H_full_2}. Our strategy will be to organize perturbation theory for disorder-averaged quantities such as the Green's function and conductivity in terms of the last two terms in $\mathcal{H}'$ (i.e. bulk and surface disorder), while incorporating the effects of the magnetic field and the averaged confining potential non-perturbatively. Our treatment of disorder will be restricted to the self-consistent Born approximation (SCBA) \cite{Ando_SCBA,Ando_conductivity}, neglecting any diagrams with crossed impurity- and boundary-scattering lines. The contribution of this type of diagrams is known to be suppressed by a factor $\omega_c/\epsilon_F$, which is small when many Landau levels are filled.  
\section{Green's function and SCBA}
\label{sec:self_energy}
In order to calculate oscillations in the conductivity, we first need to analyze the single-particle Green's function. To this end, we adopt the approach of Sheng et al. \cite{Sheng1995}, improving and generalizing their method to a non-zero magnetic field. We begin by formulating the SCBA method which accounts for scattering induced by both the bulk impurities and the surface. The full Green's function in the coordinate basis and for a fixed realization of the bulk/surface random potentials is given by
\begin{equation}
    \mathcal{G}^R(\epsilon,r,r')=\sum_{a} \frac{\psi_a(r)\psi^*_a(r') }{\epsilon-E_a+i 0^+}\;,
\end{equation}
where $\psi_a$ and $E_a$  are the eigenfunctions and eigenenergies of the full Hamiltonian in the arbitrary basis $a$. Since we are dealing with a noninteracting model, mixing between different frequencies is absent. Consequently, it suffices to evaluate the Green's function at the Fermi energy. This simplification also holds when computing the dc conductivity at zero temperature in Section \ref{sec:cond}, as the Kubo formula restricts energies to the Fermi level. Therefore, we omit the explicit frequency dependence in the calculations below, implicitly assuming $\epsilon=\epsilon_F$, unless stated otherwise.

We now calculate the disorder-averaged Green's function $\langle \mathcal{G}^R(\epsilon,r,r')\rangle$ (for brevity, we will omit the angle brackets in what follows). Within the self-consistent Born approximation (SCBA), this Green's function is given by the sum of all rainbow-type diagrams generated by the disorder correlators in Eqs.~\eqref{eq:v_average} and \eqref{eq:eta_average}. As a result, it satisfies the following self-consistent equation:
\begin{equation}
    \mathcal{G}^R(r_1,r_2)=\mathcal{G}^R_0(r_1,r_2)+\mathcal{G}_0^R(r_1,r) \Sigma^R_v(r,r')\mathcal{G}^R(r',r_2)+\mathcal{G}_0^R(r_1,r) \Sigma^R_\eta(r,r') \mathcal{G}^R(r',r_2)\;.
    \label{eq:dyson_r}
\end{equation}
Note that, due to the presence of the boundary, translational invariance is never fully restored even after disorder averaging and at $B=0$. We also introduced a bare Green's function $\mathcal{G}^R_0(r_1,r_2)$ corresponding to the first two terms in $\mathcal{H}'$ in Eq.~\eqref{eq:H_full_2}.  The bulk self-energy ($\Sigma^R_v$) and surface self-energy ($\Sigma^R_\eta$) are given by
\begin{equation}
  \Sigma^R_v(r,r')=v_0^2 \delta(r-r') \mathcal{G}^R(r,r) , \quad \quad  \Sigma^R_\eta(r,r')=\sum_{\sigma}\eta_0^2 V^2 \delta(r-r')\delta(z-\sigma d) \mathcal{G}^R(r,r)\;. 
  \label{eq:self_energy_r}
\end{equation}

The calculation of the full averaged Green's function $\mathcal{G}^R$ in Eq.~(\ref{eq:dyson_r}) can be naturally divided into two parts. First, we compute the intermediate Green's function $\mathcal{G}_v^R$ in the presence of only bulk impurities. Then we obtain the full Green's function $\mathcal{G}^R$ by adding boundary scattering. In matrix form Eq.~(\ref{eq:dyson_r}) can alternatively be rewritten as
\begin{equation}
    \begin{dcases}
            & \mathcal{G}_{v}= \mathcal{G}_0+   \mathcal{G}_0 \Sigma_{v} \mathcal{G}_{v}\\
        & \mathcal{G}= \mathcal{G}_{v}+   \mathcal{G}_{v} \Sigma_{\eta}  \mathcal{G}.\\
    \end{dcases}
    \label{eq:dyson_separated}
\end{equation}

We first find the Green's function without magnetic field ($B=0$), outlining the crucial steps of the computation and major assumptions. We then generalize the approach to the case of a nonzero magnetic field ($B\neq0$). 

\subsection{Revisiting the zero magnetic field limit ($B=0$)}
In the case of zero magnetic field, the system is translationally invariant in the $xy$-plane. Therefore, it is convenient to work in the Fourier basis. 
To this end, we express the bare Green's function as:
\begin{equation}
    \mathcal{G}^R_0(r,r')=\frac{1}{2\pi}\int d^2r_{\parallel} d^2 r_{\parallel}'e^{-i k_{\parallel}(r_\parallel-r'_\parallel)}G_0^R(k_{\parallel},z,z').
\end{equation}

The translational invariance in the $z$-direction is broken by the boundaries. In the limit $V\rightarrow \infty$ the wave functions are $\psi_\lambda(z)=\sqrt{2/d} \sin(q_\lambda z)$, where $q_\lambda= \pi \lambda/d, \quad \lambda=1,2,...$ are the discrete values of momenta in the $z$-direction. Thus, the Green's function takes the form

\begin{equation}\label{eq:G_bare_B=0}
   G^R_0(k_\parallel,z,z')=\frac{2}{d} \sum_{\lambda} \frac{\sin(q_\lambda z) \sin(q_\lambda z' ) }{k_F^2/2m-k_{\parallel}^2/2m-q_\lambda^2/2m+i 0^+}\;.
\end{equation}

Equation (\ref{eq:self_energy_r}) for the self-energy in the new basis then becomes:
\begin{equation}
    \Sigma_v^R(z,z')=v_0^2 \delta(z-z') \int \frac{d^2k_{\parallel}}{(2\pi)^2}G^R(k_\parallel,z,z).
    \label{eq:self_energy_B_0}
\end{equation}
According to our general strategy outlined in Eq.~\eqref{eq:dyson_separated}, we begin by replacing the full Green's function $G^R$ with the bulk Green's function $G^R_v$, thus neglecting the impact of boundary scattering on the bulk self-energy. This simplifies the analysis, and later we use the full Green's function to account for boundary effects. We also note that the full Green's function $ G^R_0(k_\parallel,z,z')$ is still diagonal in the in-plane momentum basis, and the self-energy is independent of the in-plane momentum due to the white-noise form of the correlation function in Eq.~\eqref{eq:v_average}. However, its dependence on the spatial coordinate $z$ cannot be neglected. We also note that the real part of the self-energy formally diverges, but could be regularized by introducing a UV cutoff associated with a finite impurity size. Its constant part could be absorbed into the redefinition of the chemical potential, and the remaining $z$-dependent part is small and rapidly oscillates away from the boundary provided that $k_F d\gg 1$, so it can be ignored. In what follows, we focus on the imaginary part of the self-energy, which is directly related to the local density of states $\Im \Sigma_v^R(z,z')=-\pi v_0^2\delta(z-z') \nu(z)$.

The problem of finding the Green's function $G_{v}(k_{\parallel},z,z')$ in this approximation is then equivalent to solving a time-independent Schrodinger equation in the complex potential $-i \pi v_0^2 \nu(z)$, where the imaginary part comes from the $z$-dependent broadening of the local density of states \cite{ZhangButler}:
\begin{equation}
 \left(   \frac{k_F^2-k_{\parallel}^2}{2m}+\frac{1}{2m}\frac{d^2}{  dz^2}+i \pi v_0^2 \nu(z) \right)G_{v}(k_{\parallel},z,z')=\delta(z-z').
 \label{eq:dyson_diff}
\end{equation}
 Solving this type of equation with the self-consistent complex potential $\nu(z)$ could be quite complicated. However, we note that the wavelength of the oscillatory term of $\nu(z)$ is $\lambda_F=\pi/k_F$ and the amplitude of the oscillations is suppressed by a factor $1/(k_F d)$ compared to density of states in the three-dimensional bulk, $\nu_0$. Since $\lambda_F$ is the shortest length-scale in the system, and $k_F d \gg 1$, we can safely average over these oscillations (see more details in the Appendix \ref{app:spat_osc}). The expression for the self-energy in this approximation becomes
\begin{equation}
    \Sigma_v^R(z,z')=v_0^2 \delta(z-z') \int \frac{d^2k_{\parallel}}{(2\pi)^2}  \overline{G^R(k_\parallel,z,z)} = \overline{\Sigma}_v^R \delta(z-z'),
\label{eq:self_en_aver}
\end{equation}
where $\overline{X} = d^{-1}\int_0^d dz X$ stands for integration over $z$.
Once the self-energy depends on $z$ only via the delta function, Dyson's equation acquires a simple form:

\begin{equation}
      G^R_v(k_\parallel,z,z')=\frac{2}{d} \sum_{\lambda} \frac{\sin(q_\lambda z) \sin(q_\lambda z' ) }{k_F^2/2m-k_{\parallel}^2/2m-q_\lambda^2/2m- \overline{\Sigma}_v^R},\quad\quad \Im \overline{\Sigma}^R_v\equiv -\frac{1}{2\tau}
      \label{eq:GR_0}
\end{equation}
As discussed previously, the real part of the self-energy renormalizes the chemical potential, while the imaginary part  $\Im \overline{\Sigma}^R_v\equiv -1/2\tau$ captures the broadening of the energy levels by impurities ($\tau$ is the mean free time). After performing a sum over $\lambda$ (or directly solving the differential equation \eqref{eq:dyson_diff}) we obtain
\begin{equation}
	G_v(k_{\parallel},z,z')=\begin{dcases}
		& \frac{2m}{\kappa}\frac{\sin(\kappa z) \sin \kappa (z'-d)}{ \sin \kappa d}, \quad z<z'\\
		&\frac{2m}{\kappa} \frac{\sin(\kappa z') \sin \kappa (z-d)}{\sin \kappa d} , \quad z>z'\\
	\end{dcases}, \quad \kappa^2=k_F^2-k_{\parallel}^2+i m/\tau.
\label{eq:Gzz1}
\end{equation}

Replacing $\overline{\sin(q_{\lambda }z)^2 }=1/2$ the expression for the self-energy after averaging in Eq.~(\ref{eq:self_en_aver}) leads to the following closed-form equation for the scattering rate $1/\tau$

\begin{equation}   \label{eq:Gamma_definition}
\frac{1}{\tau}=-\frac{2  v_0^2}{d}  \int \frac{d^2k_{\parallel}}{(2\pi)^2} \sum_{\lambda} \Im\frac{1 }{k_F^2/2m-k_{\parallel}^2/2m-q_\lambda^2/2m+ \frac{i}{2\tau}}.
\end{equation}
This approach allows us to successfully interpolate between the three-dimensional (3D) regime, where the sum over $\lambda$ can be replaced by an integral over $k_z$, and a two-dimensional (2D) regime, where only integration over $d^2k_{\parallel}$ remains. The integral can be calculated assuming $\tau \epsilon_F \gg 1$, giving $1/\tau=m v_0^2 \lambda_{\rm max}/d$, where $\lambda_{\rm max}=[k_F d/\pi]$. In 2D it simplifies to $1/\tau_{2D}=m v_0^2/d$, and in 3D it becomes $1/\tau_{3D}=m v_0^2 k_F/\pi$.

Next, we determine the effect of the boundary disorder. Following similar steps and focusing on the imaginary part, we obtain the following self-consistency equation for the boundary self-energy:
\begin{equation}
    \Sigma_\eta(z,z')= i V^2 \eta_0^2 \delta(z-z')\sum\limits_{\sigma=0,1} \delta(z-\sigma d)  \int \frac{d^2 k_{\parallel}}{(2\pi)^2}   \Im G(k_\parallel,\sigma d,\sigma d)=-i  \delta V \sum_{\sigma=0,1} \delta(z-\sigma d) \delta(z-z').
\end{equation}
where $G(k_{\parallel},0,0)=G(k_{\parallel},d,d)$ by symmetry, and thus we define a single effective scattering length $\delta$ as
\begin{equation}\label{eq:def_delta}
     \delta \equiv - \eta_0^2 V \int \frac{d^2 k_{\parallel}}{(2\pi)^2} \Im G(k_{\parallel},0,0)\;. 
\end{equation}

Physically, $\delta$ determines how diffusive the scattering from the boundary is. In the limit $\delta =0$, the boundary is perfectly flat and all scattering becomes purely specular. We also note that this quantity was denoted as $Q$ in Sheng et al. \cite{Sheng1995}

The bare Green’s function in Eq.~\eqref{eq:G_bare_B=0} vanishes identically at the boundaries and therefore cannot be used directly in Eq.~\eqref{eq:def_delta}. This issue arises because, in deriving Eq.~\eqref{eq:G_bare_B=0}, we took the limit $V \to \infty$ before incorporating the boundary disorder potential. However, since the disorder term itself is accompanied by a factor of $V$ (see Eq.~\eqref{eq:def_delta}), this results in an ill-defined “$\infty\times 0$” ambiguity. To resolve this, we return to the case of a finite square well with large but finite potential $V$, where the wave functions at the boundaries satisfy the following conditions:
\begin{equation}
    \psi_\lambda(0) V^{1/2}=(-1)^{\lambda-1} \psi_\lambda(d) V^{1/2}=\frac{q_\lambda}{\sqrt{m d}}.
\end{equation}
Using Eq.~(\ref{eq:GR_0}) with modified wave functions at the boundaries, we arrive at
\begin{equation} 
\begin{split}
 &V  G_v^R(k_{\parallel},0,0) \rightarrow \frac{2}{d}\sum_\lambda\frac{q_\lambda^2}{\kappa^2-q_\lambda^2}=\frac{1}{2m} \partial_z \partial_{z'} G_{v}^R(k_\parallel,z,z')\bigg|_{z=z'=0}=\kappa \cot \kappa d\;,\\
&V   G_v^R(k_{\parallel},0,d)\rightarrow\frac{2}{d} \sum_\lambda\frac{(-1)^{\lambda-1}q_\lambda^2}{\kappa^2-q_\lambda^2} =-\frac{1}{2m} \partial_z \partial_{z'} G_{v}^R(k_\parallel,z,z')\bigg|_{z=0,z'=d}=- \frac{\kappa}{\sin \kappa d}\;,   \\ \label{eq:G00Vlimit}
\end{split}
\end{equation}
where we evaluated the derivatives using Eq.~\eqref{eq:Gzz1}. Strictly speaking, the first term in Eq.~\eqref{eq:G00Vlimit} includes an additional contribution of order $\sqrt{V}$, which becomes evident when employing a finite-$V$ regularization. This extra term, however, is a purely real constant independent of any quantum numbers. As in the case of bulk impurities, the appearance of this divergent constant reflects a limitation of modeling the boundary disorder as delta-correlated. Consequently, it should be absorbed into a redefinition of the chemical potential. In practical terms, this regularization corresponds to replacing $V  G(k_{\parallel},0,0)$ with $\lim\limits_{z=z'\rightarrow 0}\frac{1}{2m} \partial_z \partial_{z'} G^R(k_\parallel,z,z')$. It is also worth noting that this divergence arises only for the Green’s function evaluated at the same boundary; other quantities, such as $G(k_{\parallel},0,d)$, remain well-defined and can be computed directly.

Proceeding in a similar fashion, we also find
\begin{equation}
   \begin{split}
 & V^{1/2}   G_v^R(k_{\parallel},z,0)=\sqrt{2 m } \frac{\sin \kappa (z-d)}{\sin \kappa d}\;,   \\
 & V^{1/2}   G_v^R(k_{\parallel},z,d)=-\sqrt{2 m } \frac{\sin \kappa z}{\sin \kappa d} \;.  \\
   \end{split} \label{eq:G_v_boundary}
\end{equation}

These identities allow us to solve the Dyson's equation
\begin{equation}
    G(k_{\parallel},z,z')=G_v(k_{\parallel},z,z')-i \delta V \sum\limits_{\sigma=0,1} G_v(k_{\parallel},z,\sigma d)G(k_{\parallel},\sigma d,z')\label{eq:Dyson_G_b}
\end{equation}
 in the $V\rightarrow \infty$ limit. To this end, we first set $(z,z')$ to $(0,0)$ and $(d,0)$ and solve a system of two coupled linear equations. As a result, we find 
 \begin{equation}
VG(k_{\parallel},0,0)=\frac{\kappa \cot  (\kappa d)- i \delta \kappa^2}{1+2 i \delta \kappa \cot(\kappa d)+\delta^2 \kappa^2}\;,
    \quad     VG(k_{\parallel},0,d)=\frac{\kappa \csc  (\kappa d)}{1+2 i \delta \kappa \cot(\kappa d)+\delta^2 \kappa^2}
    \;.
    \label{eq:G_on_boundary}
\end{equation}
Using these expressions in combination with Eq.~\eqref{eq:G_v_boundary}, we can then solve for the case when only one of the arguments in Eq.~\eqref{eq:Dyson_G_b} is on the boundary, while the remaining one is in the bulk. Finally, we can consider both arguments in the bulk, $0<z,z'<d$, and the solution reads as
\begin{equation}
    G^R(k_\parallel,z,z')=\frac{m}{i \kappa}\frac{\Upsilon(k_{\parallel},z,z')}{(1- R T(2d))}, \quad \quad \Upsilon(k_{\parallel},z,z')=T(z_-)+R T(2d-z_-)-\sqrt{R}(T(z_+)+T(2d-z_+)),  
    \label{eq:Green_full}
    \end{equation}
 where $R$ is the reflection coefficient, associated with the surface scattering length $\delta$ and $T(z)$ is the standard transmission function given by
\begin{equation}
    R=\left(\frac{1-\kappa \delta}{1+\kappa \delta} \right)^2, \quad \quad T(z)=e^{i \kappa z }, \quad \quad z_-=|z-z'|,\quad z_+=z+z'.
    \label{eq:R_definition}
\end{equation}
We note that $\kappa$ depends on $k_\parallel$, see Eq.~\eqref{eq:Gzz1}, which makes the reflection coefficient $R$ momentum dependent. The particles coming parallel to the surface reflect almost specularly, while particles coming perpendicular to the surface diffuse the most. By contrast, the reflection coefficient $R$ is introduced phenomenologically in the semiclassical theory and thus does not distinguish between the different particle trajectories. The 
$\Upsilon(k_{\parallel},z,z')$ describes the effective propagation between the points $z$ and $z'$, including reflections from the boundary and the corresponding interference between the trajectories.

In Appendix \ref{app:spat_osc} we show that the local density of states evaluated by plugging the full Green's function from Eq.~\eqref{eq:Green_full} into Eq.~\eqref{eq:self_energy_B_0} remains essentially unaffected by the presence of boundary disorder: the oscillations retain the same wave vector $2k_F$ and their amplitude is further suppressed. Therefore, we conclude that the averaging procedure in Eq.~(\ref{eq:self_en_aver}) still applies.

We note that the value at the boundary $V  G(k_{\parallel},0,0)$ obtained in Eq.~\eqref{eq:G_on_boundary} coincides with $\lim\limits_{z=z'\rightarrow 0}\frac{1}{2m} \partial_z \partial_{z'} G^R(k_\parallel,z,z')$ evaluated using Eq.~\eqref{eq:Green_full}. This reinforces our expectation that the regularization procedure discussed below Eq.~\eqref{eq:G00Vlimit} is self-consistent. Substituting the resulting expression into the definition of the boundary scattering length $\delta$ in Eq.~\eqref{eq:def_delta}, we obtain the following expression
\begin{equation}
    \delta= \eta_0^2 \Im \int^{k_F}_0 \frac{k_\parallel dk_\parallel}{2\pi}\frac{i \kappa}{1+\delta \kappa} \frac{1+\sqrt{R}e^{2 i \kappa d}}{1-R e^{ 2 i \kappa d}}\;.
     \label{eq:self_cons_Q}
\end{equation}
In practice, however, the self-consistent solution does not differ significantly from the first-order result obtained by using $G_v(k_{\parallel},0,0)$ in place of the full Green's function in Eq.~\eqref{eq:def_delta}, provided that surface disorder is weak.


\subsection{Extension to the finite magnetic field case ($B \neq 0$)}\label{sec:Sigma_finite_B}
Having outlined the key steps in the calculation of the self-energy and Green's function, we now consider the case of a finite magnetic field. The wave functions in the $xy$-plane become Landau states instead of plane waves, and the spectrum is $E_a=\omega_c(n+1/2)+q_\lambda^2/2m$. The bare Green's function is diagonal in the Landau basis and given by
\begin{equation}
   G^R_0(n,z,z')=\frac{2}{d} \sum_{\lambda} \frac{\sin(q_\lambda z) \sin(q_\lambda z' ) }{k_F^2/2m-\omega_c(n+1/2)-q_\lambda^2/2m+i 0^+},
\end{equation}
where $n$ denotes the Landau level. In the case of a delta potential, the expression for bulk self-energy becomes particularly simple, and it does not depend on the Landau level index \cite{Ando_SCBA}:

\begin{equation}
    \Sigma^R_v(z,z')=  \frac{v_0^2}{2\pi l_B^2}\delta(z-z') \sum_n \overline{G^R(n,z,z)} = \overline{\Sigma}_v^R \delta(z-z'),
    \label{eq:sigma_self_cons}
\end{equation}
where we average our answer over fast fluctuations on the scale of the Fermi length, as explained in the previous subsection.
Similarly, the bulk Green's function without boundary scattering ($\delta=0$) is given by Dyson's equation:

\begin{equation}
      G_v^R(n,z,z')=\frac{2}{d} \sum_{\lambda} \frac{\sin(q_\lambda z) \sin(q_\lambda z' ) }{k_F^2/2m-\omega_c(n+1/2)-q_\lambda^2/2m-\overline{\Sigma}_v^R}.
      \label{eq:GR_0_n}
\end{equation}
The self-consistency equation for the bulk self-energy then becomes
\begin{equation}   
\overline{\Sigma}_v^R=\frac{v_0^2}{d} \frac{1}{2\pi l_B^2} \sum_n  \sum_{\lambda} \frac{1 }{k_F^2/2m-\omega_c(n+1/2)-q_\lambda^2/2m-\overline{\Sigma}_v^R}.
\label{eq:self_en_full}
\end{equation}
where $l_B=1/\sqrt{B}$ is the magnetic length. In the limit of a weak magnetic field $\omega_c \tau\ll 1$, one can use the Poisson summation formula. For arbitrary thickness $d$, the expression for self-energy is

\begin{equation}
  \overline{\Sigma}^R_v= - \frac{i}{2\tau}\bigg\{1+ \frac{2}{ \lambda_{\rm max}}\sum_{s=1}^{\infty}(-1)^s R_D^s \sum_{\lambda=1}^{\lambda_{\rm max}} e^{\frac{ \pi i s}{m\omega_c}(k_F^2-q_\lambda^2)}\bigg\}, \quad R_D=e^{-\frac{\pi}{\omega_c\tau}},
   \label{eq:self_energy_Q_0}
\end{equation}
where $1/\tau$ is obtained from Eq.~\eqref{eq:Gamma_definition}, and $R_D$ denotes the standard Dingle factor \cite{Ando_SCBA} encoding damping of oscillations due to impurity scattering, and $\lambda_{\rm max}=[k_F d/\pi]$ denotes the number of modes in the $z$-direction.
The summation over $\lambda$ can be also simplified by the Poisson summation formula when $\lambda_{\rm max}\gg 1$:

\begin{equation}
 \overline{\Sigma}^R_v= - \frac{i}{2\tau}- i  \frac{ d}{\pi \tau \lambda_{\rm max}}\sum_{s=1}^{\infty} (-1)^s R_D^s \sum_{p=0}^{\infty}(2-\delta_{p,0})\int_0^{k_F} dk_z e^{\frac{ \pi i s}{m\omega_c}(k_F^2-k_z^2)}\Re e^{2 i p d k_z}.
 \label{eqn:self_energy_interm}
 \end{equation}

The integral oscillates rapidly if there are many filled Landau levels: $\epsilon_F\gg \omega_c$. Therefore, the saddle point approximation can be applied with the saddle point located at $k_{z,s}=B p d/\pi s$. It lies within the integration range only if $p< p_s=\pi k_F s/Bd$. For $p>p_s$, no saddle point falls inside the interval $[0,k_F]$. The integral must instead be approximated by boundary contributions obtained via integration by parts. These boundary terms are non-oscillatory and smaller by a factor of $k_F l_B$ compared to the saddle-point contribution. As a result, higher-order harmonics with $p>p_s$ are strongly suppressed and do not contribute to the oscillations; they can be safely neglected by imposing a hard cutoff in the sum over $p$. When $s$-harmonics are suppressed by a Dingle factor, $p$-harmonics are similarly indirectly suppressed due to the effective cutoff $p_s$. Finally, we obtain
 \begin{equation}
 \overline{\Sigma}^R_v=- \frac{i}{2 \tau}-  \frac{ id/l_B}{2 \pi \tau \lambda_{\rm max} }
  \sum_{s=1}^{\infty} \frac{
(-1)^s}{\sqrt{s}}   R_D^s \sum_{p=0}^{p_s}(2-\delta_{p,0})\exp \left(2 \pi i s \frac{\epsilon_F}{\omega_c} +  \frac{ip^2 }{\pi s} \left(\frac{d}{l_B}\right)^2 -\frac{i\pi}{4}\right).
\label{eqn:self_energy_full}
\end{equation}

If we only retain the $p=0$ harmonic, then this expression becomes identical to the standard 3D quantum oscillations. On the other hand, to recover the 2D limit, one can take Eq.~\eqref{eq:self_energy_Q_0} and remove the sum over $\lambda$ altogether. When $s \gtrsim s_{\rm max}=\omega_c \tau/\pi$ the contribution from all modes $s$ is negligible, due to exponential suppression by the Dingle factor $R_D$. Consequently, through the cutoff $p<p_s$, the maximum allowed value of $p$ is $p_{\rm max}=l/d$. The result is intuitively clear: all harmonics associated with the finite-size effects are suppressed once $d\gg l$, and in the opposite limit $d\ll l$ many discrete harmonics become relevant. Fig.~\ref{fig:self_energy_osc} (a),(b) shows quantum oscillations of the self-energy in two regimes $d/l=5$ and $d/l=0.5$. The first regime is close to the 3D case, where the amplitude of the oscillations is suppressed by the factor $\sqrt{\omega_c/\epsilon_F}$. Moreover, since $\omega_c \tau \sim 1$ only a few harmonics contribute significantly. In contrast, the second regime is close to the quasi-2D case since $d/l<1$. It exhibits more complex oscillations due to strong size effects, which are accurately reproduced by Eq.~(\ref{eqn:self_energy_full}).





\begin{figure}[h]
      \begin{minipage}[h]{1\linewidth}
    \center{\includegraphics[width=0.45\linewidth]{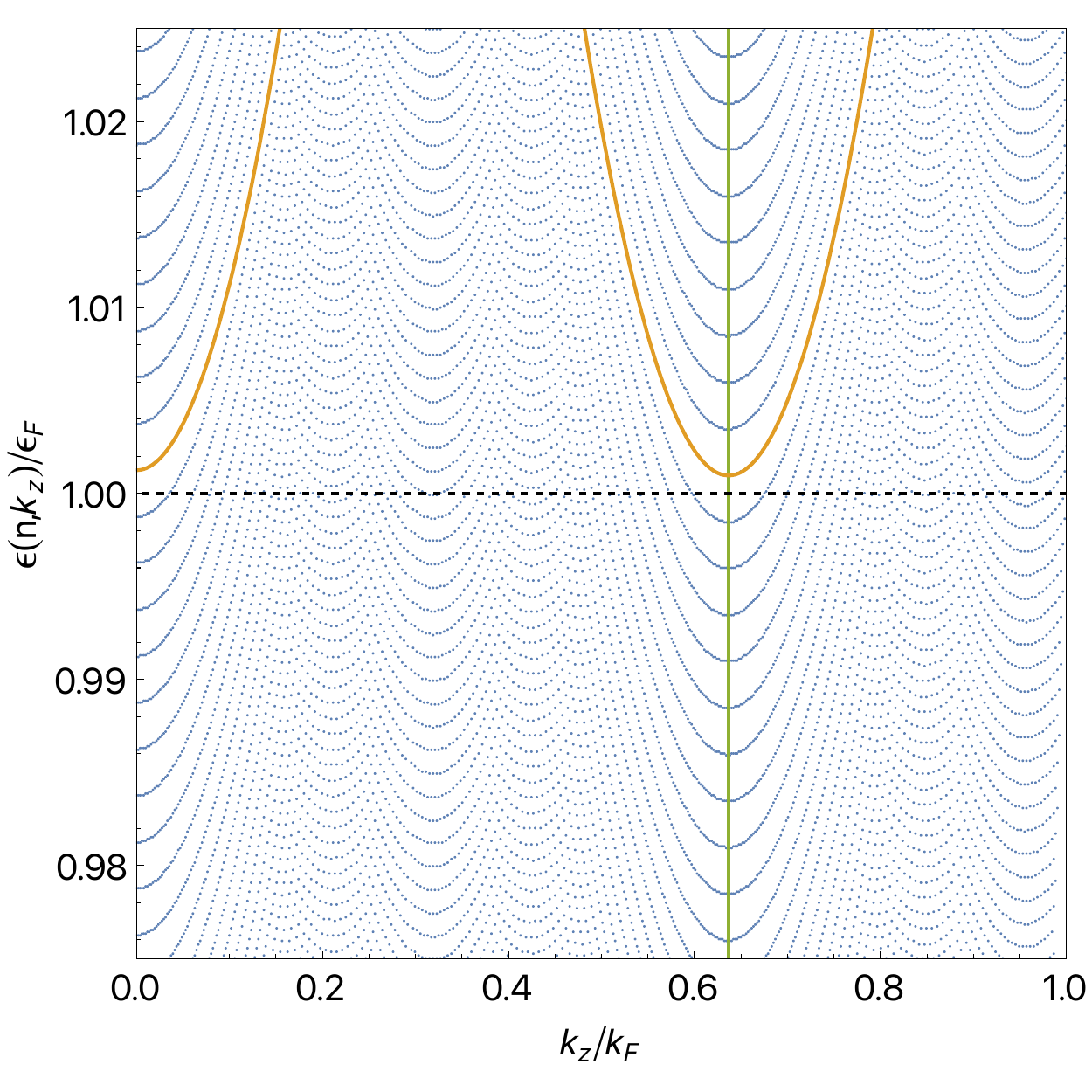}}
    \end{minipage} 
\caption{Discrete energy levels from Eq.~(\ref{eq:dispersion}) for parameters $\omega_c/\epsilon_F=1/400$, $k_F d=1600$. 
The black dashed line denotes the chemical potential at the Fermi energy. Blue dots denote the individual energy levels. Due to the discreteness, the spectrum can be partitioned into set of parabolas (orange lines) shifted from $k_z=0$, see the discussion in the main text. The green line shows $\bar{k}_z$ with $p=s=1$. }
\label{fig:s_p_1_bands}
\end{figure}
\begin{figure}[h]
    \begin{minipage}[h]{0.3\linewidth}
    \center{\includegraphics[width=1\linewidth]{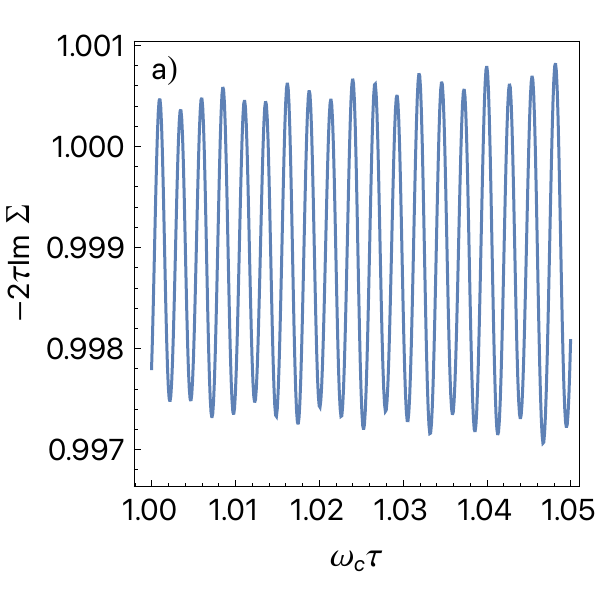}}
    \end{minipage} 
    \hfill
    \begin{minipage}[h]{0.3\linewidth}
    \center{\includegraphics[width=1\linewidth]{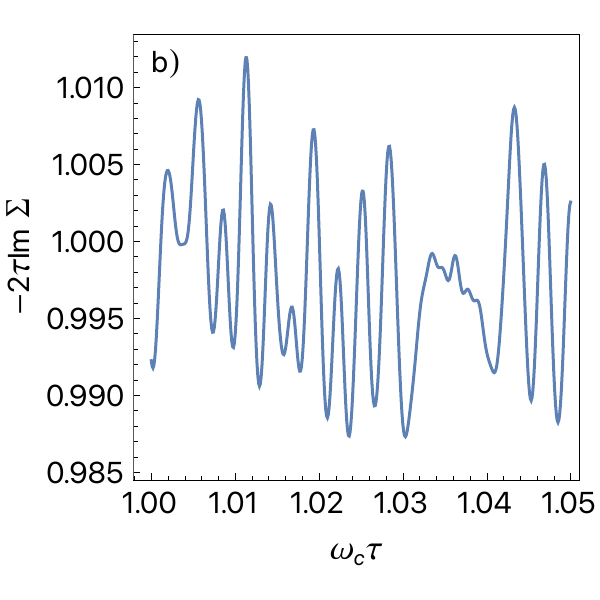}}
    \end{minipage} 
    \hfill
        \begin{minipage}[h]{0.3\linewidth}
    \center{\includegraphics[width=1\linewidth]{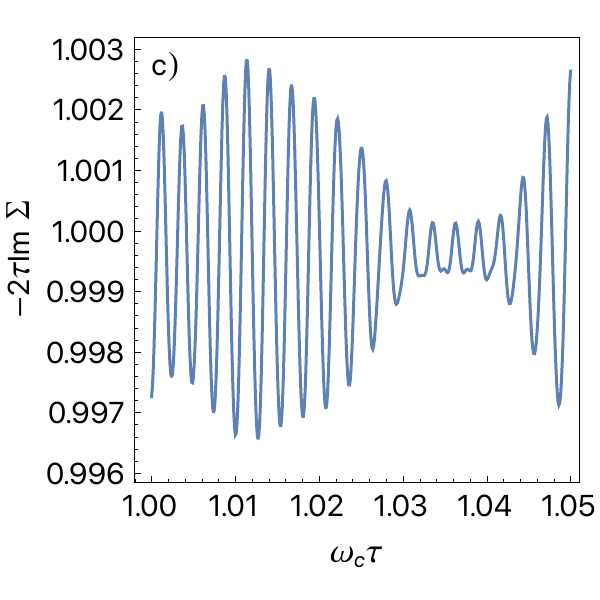}}
    \end{minipage} 
\caption{Imaginary oscillating part of the self-energy.
(a) $\delta=0$, $k_F d=4000$, $d/l=5$ (3D case), (b) $\delta=0$, $k_F d=400$, $d/l=0.5$ (quasi 2D case, specular scattering), (c) $\delta=0.5/k_F$,  $k_F d=400$, $d/l=0.5$ (quasi 2D case, diffusive boundary scattering).}
\label{fig:self_energy_osc}
\end{figure}

Various frequencies in Eq.~(\ref{eqn:self_energy_full}) could also be captured from the underlying dispersion relations, as was noted in \cite{Taen2023}. In the continuum limit, peaks in the density of states occur when the bottom of the $k_z$-dependent band crosses the Fermi energy, which happens at $n=\epsilon_F/\omega_c$. In the discrete case, in addition to the usual parabolic bands centered at $k_z=0$, there are additional parabolas centered at $\bar{k}_z=B p d/(\pi s)$, see Fig.~\ref{fig:s_p_1_bands}. These values of $\bar{k}_z$ coincide with the saddle points of the integral in Eq.~(\ref{eqn:self_energy_interm}), which shows a deeper connection between the two approaches. To derive the expression for  $\bar{k}_z$, consider the general discrete spectrum and choose the discrete values of $k_z,n$ according to:
\begin{equation}
\epsilon_{n,k_z}=\frac{k_z^2}{2m}+\omega_c(n+1/2), \quad k_z=\frac{\pi (s \lambda+j)}{d}, \quad n=c-p \lambda , \quad \quad \lambda \in \mathds{Z}, j=0,..s-1. 
\label{eq:dispersion}
\end{equation}
Using simple manipulations, we can alternatively write:
\begin{equation}
\epsilon_{n,k_z}=\frac{1}{2m}\left(\left(\frac{\pi (s \lambda+j)}{d} \right)^2-2\bar{k}_z\left(\frac{\pi (s \lambda+j)}{d} \right) \right)+\omega_c\left(c+\frac{1}{2}+\frac{p j}{s} \right).
\end{equation}
Therefore, our discrete points belong to parabolas centered at $\bar{k}_z$:
\begin{equation}
\epsilon_{n,k_z}=\frac{1}{2m}\left(k_z-\bar{k}_z\right)^2+\omega_c\left(c+\frac{1}{2}+\frac{p j}{s}\right)+\frac{\bar{k}_z^2}{2m}
\end{equation}
Let us introduce $\tilde{c}=c+p j/s$ as a new Landau levels index. 
We note that as $s$ increases, the Landau levels become more closely spaced, since the index $\tilde{c}$ can take fractional values with denominator $s$, i.e., $\tilde{c}  \in \mathds{Z}/s$.
The parabolic bands cross the Fermi level when $\epsilon_F=\omega_c (\tilde{c}+1/2)+\bar{k}_z^2/(2m)$. Each time such a crossing occurs, we expect a resonance in the density of states and in the self-energy, meaning they oscillate as follows $\Sigma \propto \cos (2 \pi s \tilde{c})$. Substituting the definition of $\tilde{c}$ we recover the oscillatory behavior: $\Sigma\propto \cos (2 \pi s \epsilon_F/\omega_c+B p^2 d^2/(\pi s))$, with the frequencies matching Eq.~(\ref{eqn:self_energy_full}).

There is another elegant quasiclassical argument that supports Eq.~(\ref{eqn:self_energy_full}). Consider particles that return to their initial positions after reflecting off the two surfaces $p$ times. The $k_z$ component of their momenta is quantized according to $2\omega_c pd m/\bar{k}_z=2 \pi s$,which corresponds to the particle completing $s$ full rotations. Solving for the momentum gives $\bar{k}_z=B p d/(\pi s)$. The semiclassical trajectories in the momentum space are orbits shifted from the equator with the enclosed area $\mathcal{S}=\pi(k_F^2-\bar{k}_z^2)$. Using the Bohr quantization condition $\mathcal{S}/B=2 \pi n$ we reproduce the frequencies derived earlier. From the real-space quantization viewpoint, the factor $d^2/l_B^2$ that appears in the frequency represents the number of magnetic-flux quanta threading a square of side length $d$. The hard cutoff $p<p_s$ could also be explained quasiclassically in a direct way. For electrons with $p>p_s$ the accumulated phase after $p$ reflections from both surfaces exceeds $(2 d p_s/v_F) \omega_c=2\pi s$. Therefore, for such $(p,s)$ pairs no resonance condition is satisfied.


So far, we have analyzed finite-size effects in the oscillatory self-energy for purely specular boundary scattering. Discreteness of the momentum states in the $z$-direction resulted in the appearance of a second sum over harmonics in Eq.~\eqref{eqn:self_energy_full}, accompanied by the phase shifts that are linear in the magnetic field. Next, we analyze the effect of surface disorder. Upon incorporating the resulting surface scattering, the expression for the full Green's function remains similar to that in the zero magnetic field case, see Eqs. (\ref{eq:Green_full}) and (\ref{eq:R_definition}):
\begin{equation}
    G^R(n,z,z')=\frac{m}{i \kappa}\frac{\Upsilon(n,z,z')}{(1- R T(2d))}, \quad  \kappa^2=k_F^2-2 m \omega_c(n+1/2)-2 m \overline{\Sigma}_v^R\;,
    \label{eq:full_G_with_boundary_finite_B}
    \end{equation}
with the reflection coefficient $R=(1-\kappa \delta)^2/(1+\kappa \delta)^2$, the transmission function $T(z)=e^{i \kappa z}$ and the effective propagator $\Upsilon(n,z,z')$ introduced earlier in Eq.~(\ref{eq:Green_full}) as a linear combination of several interfering trajectories. There are two important modifications compared to the $B=0$ case: the parallel momentum basis $k_\parallel$ is replaced with the Landau basis $n$, and $\overline{\Sigma}_v^R$ is now the self-energy in a magnetic field. Our next step is to compute the effect of the boundary scattering on $\overline{\Sigma}_v^R$ and $\delta$. To this end, we use Eq.~(\ref{eq:sigma_self_cons}):
\begin{equation}
 \overline{\Sigma}_v^R =\frac{\omega_cd}{2\pi  \lambda_{\rm max} \tau} \sum_n \overline{G^R(n,z,z)} ,
\end{equation}
where $G^R(n,z,z)$ is taken from Eq.~(\ref{eq:full_G_with_boundary_finite_B}).
After performing the integration over the $z$-direction and applying the Poisson summation formula, we obtain 
\begin{equation}
   \overline{\Sigma}_v^R =\frac{\omega_cd}{2\pi  \lambda_{\rm max} \tau} \sum_{s=0}^{\infty} \int dn \frac{m }{i \kappa}\left( 1+Re^{2i \kappa d}+i \sqrt{R}\frac{e^{2 i \kappa d}-1}{\kappa d}\right) \frac{e^{2 \pi i  s n}}{1- R e^{2 i \kappa d}} .
   \label{eq:self_energy_poisson}
\end{equation}
Next, we change the integration variable from $n$ to $\kappa$ and then use the residue theorem to evaluate the remaining integral. The poles of the integrand are located at
\begin{equation}
   \left(\frac{1-\kappa \delta}{1+\kappa \delta} \right)^2 e^{2 i \kappa d}=1 \Rightarrow \kappa \approx q_\lambda+i \kappa_2,  \quad \kappa_2=\frac{1}{d}\log   \left(\frac{1-q_\lambda \delta}{1+q_\lambda \delta} \right), \quad n=\frac{k_F^2-\kappa^2-2m\overline{\Sigma}_v^R}{2 B} \pm \frac{1}{2},
   \label{eq:poles}
\end{equation}
where as before $q_\lambda=\pi \lambda/d$. Evaluating the residues at these poles yields the following expression for the self-energy:
\begin{equation}\label{eq:Sigma_v_Bfinite_1}
     \overline{\Sigma}_v^R=- \frac{i }{2\tau}- \frac{i}{  \tau\lambda_{\rm max}}\sum_{s=1}^{\infty}  \sum_{\lambda=1}^{\lambda_{\rm max}}\left[ 1+ \frac{i(1-\tilde{R})}{2\pi \lambda \sqrt{\tilde{R}}} \right]  \exp \left(2 \pi i s \left( \frac{k_F^2-(q_\lambda+i \kappa_2)^2-2 m \overline{\Sigma}_v^R}{2 B} - \frac{1}{2}\right)\right) ,
\end{equation}
where $\tilde{R}= R|_{\kappa \rightarrow q_\lambda} = (1-q_\lambda \delta)^2/(1+q_\lambda \delta)^2$.
The second term in the square brackets can be neglected as long as  $k_F d \gg 1$. Moreover, for $\omega_c\tau\ll 1$ the oscillatory part of $\overline{\Sigma}_v^R$ on the right-hand side of Eq.~\eqref{eq:Sigma_v_Bfinite_1} can be ignored. Finally, we obtain
\begin{equation}
 \overline{\Sigma}^R_v= - \frac{i}{2\tau}\bigg\{1+ \frac{2}{ \lambda_{\rm max}}\sum_{s=1}^{\infty}(-1)^s R_D^s \sum_{\lambda=1}^{\lambda_{\rm max}} e^{\frac{ \pi i s}{m\omega_c}(k_F^2-(q_\lambda+i \kappa_2)^2)}\bigg\}
      \label{eq:self_energy_Q}
\end{equation}
The result differs from the self-energy in the specular limit by an additional shift of $q_\lambda$ by the imaginary momentum component $\kappa_2$, See Eq.~(\ref{eq:self_energy_Q_0}) for comparison.

In the limit $\lambda_{\rm max} \gg1$, $\epsilon_F \gg \omega_c$, we can once again apply the Poisson summation formula and use the saddle point approximation to evaluate the integral. The saddle point is determined by the following equation:

\begin{equation}
    k_z= \frac{B p d }{\pi s}+\frac{i}{d} \log\left(\frac{1-k_z \delta}{1+k_z \delta}\right) \quad  \Rightarrow\quad    \bar{k}_z\approx \frac{B p d }{\pi s}+\frac{i}{d} \log\left(\frac{1-\delta B p d/(\pi s)}{1+\delta B p d/(\pi s)} \right),
\end{equation}
where we assumed the imaginary part is small, and account for the shift in $k_z$ in the lowest-order. Substituting the modified saddle-point value in the integral, we find
 \begin{equation}
 \overline{\Sigma}^R_v= - \frac{i}{2\tau}-  \frac{ id/l_B}{2 \pi \tau \lambda_{\rm max}}
  \sum_{s=1}^{\infty}\frac{(-1)^s}{\sqrt{s}}
R_D^s  \sum_{p=0}^{p_s}(2-\delta_{p,0})  \left(\frac{1-\delta k_F p/p_s}{1+\delta k_F p/p_s} \right)^{2p} \exp \left(2 \pi i s \frac{\epsilon_F}{\omega_c} + i \frac{B p^2 d^2}{\pi s} -\frac{i\pi}{4}\right),
\label{self_energy:full}
\end{equation}
where as previously $p_s=\pi k_F s/Bd$. Similarly to a Dingle factor $R_D$, higher order harmonics in $p$ are additionally suppressed by a factor that depends on boundary scattering length. Fig.~\ref{fig:self_energy_osc} (c) shows oscillations in the self-energy after incorporating surface scattering, as in Eq.~(\ref{self_energy:full}). The amplitude is noticeably suppressed compared to the fully specular scattering ($\delta=0$) case.

Finally, the self-consistency equation for $\delta$, given in Eq.~\eqref{eq:def_delta}, is modified by replacing the integral over $k_\parallel$ with a summation over Landau levels. Using the Green's function from  Eq.~\eqref{eq:full_G_with_boundary_finite_B} and evaluating it at the  boundary via the relation $V  G(k_{\parallel},0,0) =\lim\limits_{z=z'\rightarrow 0}\frac{1}{2m} \partial_z \partial_{z'} G^R(k_\parallel,z,z')$, we obtain 
\begin{equation}
     \delta= \frac{\eta_0^2}{2 \pi l_B^2} \Im \sum_n  \frac{i \kappa}{1+\delta \kappa} \frac{1+\sqrt{R}e^{2 i \kappa d}}{1-R e^{ 2 i \kappa d}}\;.
\end{equation}
However, for physically relevant parameters this modification is minimal, compared to zero magnetic field case, see Eq.~\eqref{eq:self_cons_Q}. This is because quantum oscillation effects originate from the extremal orbit at the equator of the Fermi surface, while the dominant contribution to $\delta$ comes from the poles of the Fermi surface. Therefore, in the rest of this work, we neglect the oscillatory contribution to the boundary scattering length $\delta$ arising from the magnetic field.

\section{Oscillations in conductivity}
\label{sec:cond}

After determining the Green's functions in the presence of surface scattering and magnetic field, we now proceed with the calculation of the DC conductivity. Using the Kubo formula and following standard steps \cite{Bruus2004,Ando_conductivity}, we express the $T=0$ conductivity in terms of Green's functions as follows 
\begin{equation}
  \sigma^{xx}=\frac{\omega_c^2}{2\pi^2  d }\int_0^d dz dz' \sum_{n} n \;\Im G^R(n ,z,z')\Im G^R(n-1 ,z,z'). 
\label{eq:conductivity_general}
\end{equation}
where the disorder-averaged Green's functions are evaluated at the Fermi energy, and an additional factor of $n$ comes from the current matrix elements in the Landau basis \cite{Ando_conductivity,Bastin1971}. Although our primary interest lies in the case with boundary disorder, which is crucial for reproducing semiclassical Sondheimer oscillations, we begin by examining the specular scattering limit ($\delta = 0$), as it already captures quantum geometric oscillations.


\subsection{Specular boundary scattering}
In the limit of specular scattering, the Green's function $G^R_v$ is diagonal in the $q_\lambda$ basis and is given by Eq.~(\ref{eq:GR_0_n}). The expression for the conductivity (which we denote as $\sigma_v^{xx}$) simplifies to the form
\begin{equation}
    \sigma^{xx}_v=\frac{\omega_c^2}{2 \pi^2 d }\sum_{n,\lambda} n \;  \Im G^R_v(n,q_\lambda) \Im G^R_v(n-1,q_\lambda)\;.
\label{eq:conductivity_general_Q_0}
\end{equation}

 In the limit of weak magnetic field ($\omega_c \tau \ll 1$) we apply the Poisson summation formula to evaluate the sum over Landau levels, and obtain
\begin{equation}
    \sigma^{xx}_v=  \frac{\tau_B/m}{1+\omega_c^2 \tau_B^2} \sum_{\lambda=1}^{\lambda_{\rm max}} \frac{k_F^2-q_\lambda^2}{4 \pi d} \left(1+\sum_{s=1}^{\infty}2(-1)^s e^{-\frac{\pi s}{\omega_c\tau_B}}  \;\Re e^{ \pi i s \frac{k_F^2-q_\lambda^2}{ m \omega_c}  }\right)\;.
    \label{eqn:cond_zero_Q_arb_d}
\end{equation}
where $1/\tau_B \equiv -2\Im \bar{\Sigma}^R_v$ is the magnetic field dependent scattering rate, and the self-energy was derived in the previous section and is given in Eq.~\eqref{eqn:self_energy_full}. The  $s=0$ harmonic reproduces the non-oscillatory part of the Sondheimer's formula (compare to Eq.~\eqref{eq:sigma_class_finiteB} with $R=1$) with a corrected averaged particle density, provided we ignore magnetic field oscillations in $\tau_B$: 
\begin{equation}
    \sigma_{v,0}^{xx}=\frac{n \tau}{m(1+\omega_c^2 \tau^2)},\quad n=\sum_{\lambda=1}^{\lambda_{\rm max}} \frac{k_F^2-q_\lambda^2}{4\pi d}\approx \frac{k_F^3}{6 \pi^2} \left( 1-\frac{3 \pi }{4 k_F d} +\mathcal{O}((k_F d)^{-2})\right).
    \label{eq:density_quasi2d}
\end{equation}
By further setting $\omega_c\tau=0$, we recover the Drude conductivity in the absence of a magnetic field, previously denoted by $\sigma_0$. Moreover, restoring magneto-oscillations in the scattering time, $\tau \rightarrow \tau_B$, accounts for Landau quantization effects on the diffusion coefficient and the local density of states. These quantum oscillations are typically suppressed by factors of $R_D$ when $\omega_c \tau \ll 1$. Higher order harmonics in Eq.~\eqref{eqn:cond_zero_Q_arb_d} ($s\geq 1$) involve subtle mixing between geometric size effects (i.e., discreteness of $q_\lambda$) and oscillations with the magnetic field. This interplay can be made more apparent by applying the Poisson summation formula to compute the sum over $\lambda$:
\begin{equation}
 \sigma^{xx}_{v, {\rm osc}}=\frac{\tau_B/m}{  1+\omega_c^2 \tau_B^2}\sum_{s=1}^{\infty} (-1)^s  e^{-\frac{\pi s}{\omega_c\tau_B}}\sum_{p=0}^{\infty} \int_0^{k_F} \frac{dk_z}{2\pi^2}  (k_F^2-k_z^2)
\Re  \exp \left(2 \pi i s  \frac{k_F^2-k_z^2}{2m \omega_c} +2  i p d k_z \right)\;.
\end{equation}

The saddle point of this integral is located at $k_{z,s}= B p d/(\pi s)$, which was previously encountered in the self-energy calculation. If the saddle point lies above $k_F$, the contribution is negligible. Therefore, for a given $s$, only modes with $p<p_s=\pi k_F s/(B d)$ contribute. The integral can be evaluated for allowed harmonics $p,s$ using the saddle point approximation. The expression we obtain is the following:

\begin{equation}
 \sigma^{xx}_{v, {\rm osc}}=\frac{\tau_B/m}{1+\omega_c^2 \tau_B^2}\sum_{s=1}^{\infty} \frac{(-1)^s}{4\pi^2 l_B\sqrt{s}} e^{-\frac{\pi s}{\omega_c\tau_B}}  \sum_{p=0}^{p_s} (2-\delta_{p,0})  \left(k_F^2-\left(\frac{B p d}{\pi s} \right)^2 \right)
\cos \left(2 \pi  s \frac{\epsilon_F}{\omega_c} +  \frac{ p^2 d^2}{\pi s\; l_B^2} -\frac{\pi}{4}\right).
\label{eq:cond_full_simplified}
\end{equation}
Let us consider lowest-order correction to the conductivity with $p=0,1$ and $s=1$ term, provided that $ d< \pi k_F l_B^2$ and $p=1$ harmonic is present. In addition, we also expand $1/\tau_B\equiv -2\Im \bar{\Sigma}_v$ using Eq.~\eqref{self_energy:full} up to the $p=0,1$ and $s=1$ terms. After combining all factors, we find
\begin{equation}
\begin{split}
 \frac{ \sigma^{xx}_v}{\sigma^{xx}_{v,0}} &\approx  1-\frac{ (1+5 \omega_c^2 \tau^2)}{2(1+\omega_c^2 \tau^2)} R_D\sqrt{\frac{\omega_c}{2\epsilon_F}}  \cos\left(2\pi  \frac{\epsilon_F}{ \omega_c}- \frac{\pi}{4} \right)- \left(\frac{ 1+5 \omega_c^2 \tau^2}{1+\omega_c^2 \tau^2} -\frac{3}{2\pi^2} \frac{\omega_c d^2}{\epsilon_F l_B^2} \right) R_D\sqrt{\frac{\omega_c}{2\epsilon_F}} \cos\left(2\pi  \frac{\epsilon_F}{ \omega_c}+\frac{d^2}{\pi l_B^2}- \frac{\pi}{4} \right),\\
 \end{split}
 \label{eq:conductivity_quasi2d_Q_0}
\end{equation}
and $\sigma^{xx}_{v,0}$ is the non-oscillating Drude conductivity in magnetic field, given in Eq.~\eqref{eq:density_quasi2d}. The first term corresponds to the standard Shubnikov-de-Haas oscillations, while the second term oscillates with the modified frequency, which arise from quantum geometric effects. The extension of our formulas to finite temperature can be done easily by replacing Eq.~\eqref{eq:conductivity_general_Q_0} with
\begin{equation}
  \sigma^{xx}_v(T)=-\int d \omega \frac{\partial n_F(\omega)}{\partial \omega} \sigma^{xx}_v(\omega+\epsilon_F)\;,
\end{equation}
where $n_F(\omega)=1/(1+\exp(\omega/T))$ is Fermi-Dirac distribution. This leads to the additional suppression of oscillations due to thermal smearing encoded in the Lifshitz-Kosevich factor
\begin{equation}
R_T(s)= \int d \omega \frac{\partial n_F(\omega)}{\partial \omega} e^{2 \pi i s \omega/\omega_c}=\frac{2 \pi^2 s T/\omega_c }{\sinh \left(2 \pi^2 s T/\omega_c  \right)}\;.\label{eq:R_T_s}
\end{equation}
Thus, the analogue of Eq.~(\ref{eq:cond_full_simplified}) at finite temperature has the same form but with an extra factor $R_T(s)$ under the sum. All harmonics are suppressed exponentially when $T \gg \omega_c$, since the thermal energy exceeds the Landau level spacing. If $T<\omega_c$ only $s_T\sim\omega_c/T$ harmonics are left. Similarly, the $p$-harmonics are indirectly affected through the cutoff $p<p_s$. At finite temperature, only harmonics up to $p_T\sim\epsilon_F/(k_F d T)$ survive. When $T> \epsilon_F/(k_F d)$, only the $p=0$ harmonic is left and the discretization effects disappear. This reflects the fact that the thermal energy exceeds the spacing between the discrete levels in the $z$-direction.

\begin{figure}[h]
    \begin{minipage}[h]{0.3\linewidth}
    \center{\includegraphics[width=1\linewidth]{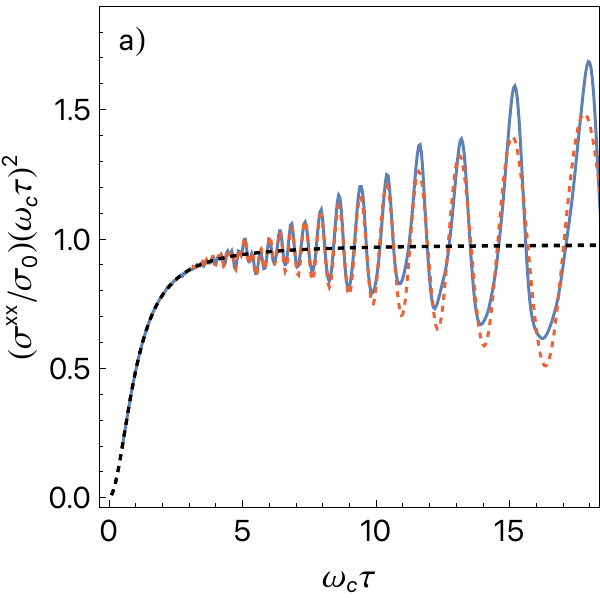}}
    \end{minipage} 
    \hfill
      \begin{minipage}[h]{0.32\linewidth}
    \center{\includegraphics[width=1\linewidth]{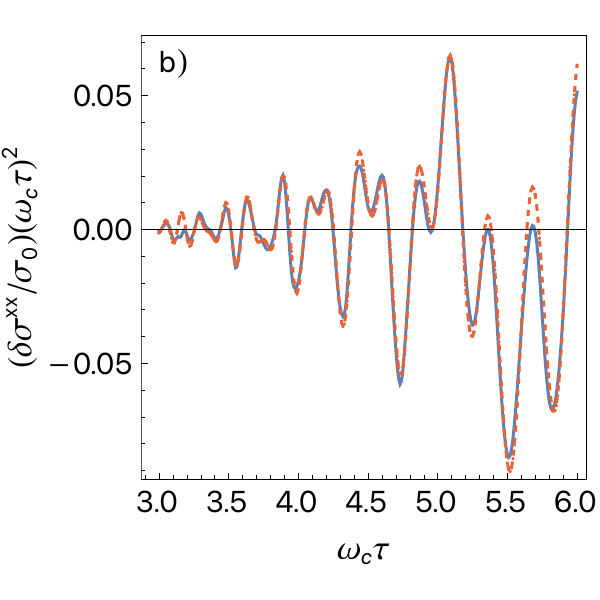}}
    \end{minipage} 
       \hfill
      \begin{minipage}[h]{0.32\linewidth}
    \center{\includegraphics[width=1\linewidth]{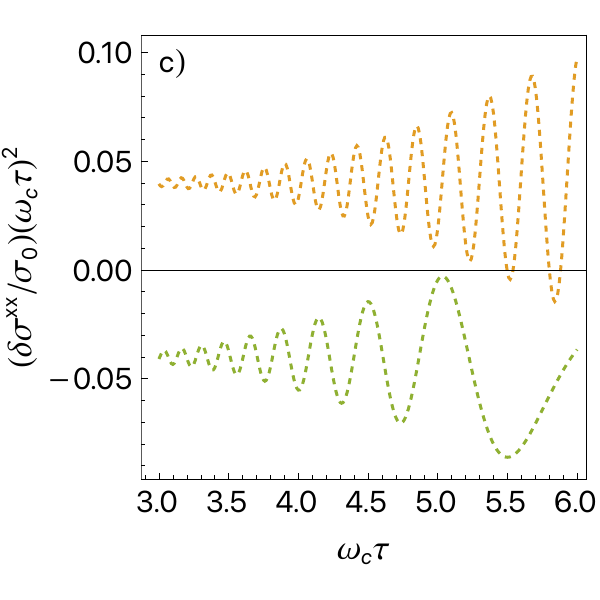}}
    \end{minipage}  
\caption{ Oscillations of conductivity in the specular scattering limit. (a) The blue line corresponds to the direct numerical evaluation of Eq.~\eqref{eq:conductivity_general_Q_0}, black dashed line represents the non-oscillating part, given by Eq.~(\ref{eq:density_quasi2d}), and the red line is the approximate formula in Eq.~\eqref{eq:conductivity_quasi2d_Q_0}. (b) Enlarged part of subfigure (a) focusing on size effects at intermediate field strength. The non-oscillatory contribution has been subtracted. (c) Two separate contributions to the oscillating part of $\sigma^{xx}$. The orange curve corresponds to the $s=1, p=0$ harmonic, whereas the green line shows $s=1,p=1$ harmonic. The parameters used are: $T/\epsilon_F=1/100$, $d/l=0.5$, $k_F d=100$, and $\delta k_F=0.5$.}
\label{fig:conduct_1}
\end{figure}

Fig.~\ref{fig:conduct_1} depicts representative quantum oscillations of the conductivity, scaled by $(\omega_c\tau)^2$, computed numerically from the full expression in Eq.~\eqref{eq:conductivity_general_Q_0} and compared against the approximate result in Eq.~\eqref{eq:conductivity_quasi2d_Q_0}. The non-oscillatory component saturates at high magnetic fields, consistent with Eq.~\eqref{eq:sigma_class_finiteB}, while the oscillatory behavior follows Eq.~\eqref{eq:cond_full_simplified} and includes the Lifshitz-Kosevich damping factor $R_D$. Fig.~\ref{fig:conduct_1}(b) shows good agreement between the full numerical result and the two-harmonic approximation in the intermediate field regime. Panel (c) further illustrates the individual contributions from each term in Eq.~\eqref{eq:conductivity_quasi2d_Q_0}, clearly revealing the presence of two distinct types of oscillations.

\subsection{Diffusive boundary scattering}
Now we compute the conductivity in the presence of boundary disorder. We use the full expression for the conductivity given in Eq.~\eqref{eq:conductivity_general} supplemented by the Green's function from Eq.~\eqref{eq:full_G_with_boundary_finite_B}. After applying the Poisson summation formula to perform the sum over Landau levels, we first single out the non-oscillatory term (i.e. the $s=0$ harmonic)
\begin{equation}\label{eq:Sigma_0_diffusive}
  \sigma^{xx}_0=\frac{\omega_c^2}{4\pi^2 d}\int dz \int dz'  \int dn n \Re \left( G^R(n,z,z')G^A(n-1,z,z') \right)\;.\end{equation}
After substituting the Green's functions from Eq.~(\ref{eq:full_G_with_boundary_finite_B}) we obtain
\begin{equation}
    \sigma^{xx}_0= \frac{d}{(2\pi l_B^2)^2}\int dn n \;\Re \left( \frac{\overline{\Upsilon(n,z,z')\Upsilon^*(n-1,z,z')}}{\kappa(n) \kappa(n-1)(1-RT(n,2d))(1-R T^*(n-1,2d))}\right),
     \end{equation}
where $\bar{X}  = d^{-2}\int_0^d \int_0^ddz dz' X(z,z')$ stands for the integrals over all $z$ coordinates. The average over $z,z'$ can be calculated exactly, reducing the problem to a single summation over Landau levels. The summation can always be performed numerically, however, to make analytical progress, we introduce several approximations. First, in the expression for $\Upsilon(n) \Upsilon^*(n-1)$, we neglect all cross terms, such as $T(z_-)T^*(z_+)$ and retain only the square terms such as $T(z_-)T^*(z_-)$. Second, we use an approximate expression for $\kappa\approx \bar{\kappa}+i m /(2 \tau \bar{\kappa})$, where $\bar{\kappa}^2=k_F^2-2 m \omega_c(n+1/2)$. Within the same level of accuracy as in Eq.~\eqref{eq:Sigma_0_diffusive}, we have also neglect the difference between $\tau_B$ and $\tau$, which arises from the higher-order $s$ harmonics. Finally, we approximate $\kappa(n) \kappa(n-1) \approx\kappa(n)^2$.
With these simplifications, we can carry out the averaging procedure and obtain the following simplified expression for the averaged value:

\begin{equation}
   \overline{\Upsilon(n,z,z')\Upsilon^*(n-1,z,z')}=\frac{2 \bar{\kappa}^2}{\zeta^2 k_F^2}\left( 1+R e^{-\zeta k_F/\bar{\kappa}}\right) \left[ (e^{-\zeta k_F/\bar{\kappa}}-1)(1-R)+\frac{\zeta k_F}{\bar{\kappa}}(1-R e^{-\zeta k_F/\bar{\kappa}})\right], 
   \label{eq:average_Gamma}
\end{equation}
where the combination $\zeta=\frac{d}{l}(1+i \omega_c \tau)$ has already appeared in Sec.~\ref{sec:sondh}. Next, we change the integration variable from $n$ to $\bar{\kappa}$:
\begin{equation}
  \sigma^{xx}_0=\frac{d}{8 \pi^2 }\int_0^{k_F} d\bar{\kappa} \frac{k_F^2-\bar{\kappa}^2}{\bar{\kappa}} \Re \left( \frac{\overline{\Upsilon(n,z,z')\Upsilon^*(n-1,z,z')}}{(1-R  e^{2i\bar{\kappa} d})(1-R  e^{-2\zeta k_F/\bar{\kappa}-2i \bar{\kappa} d})}\right)\;.\end{equation}

The integration can be performed using the residue theorem, as was done in computing the self-energy in Sec.~\ref{sec:Sigma_finite_B}. As in the previous section, the poles are located at $1-R  e^{2i \bar{\kappa} d}=0$, see Eq.~(\ref{eq:poles}) for details. The residue of the denominator is $-\pi/(d(1-\tilde{R}^2 e^{-2\zeta k_F/q_\lambda}))$, where  $\tilde{R}= R|_{\kappa \rightarrow q_\lambda} = (1-q_\lambda \delta)^2/(1+q_\lambda \delta)^2$.
Therefore, the final formula is
\begin{equation}
 \sigma^{xx}_0=\Re \frac{\tau/m}{1+i\omega_c\tau}\sum_{\lambda=1}^{\lambda_{\rm max}} \frac{k_F^2-q_\lambda^2}{4\pi d} \left(1-\frac{q_\lambda(1-\tilde{R})(1-e^{-\zeta k_F/q_\lambda})}{\zeta k_F(1-\tilde{R} e^{-\zeta k_F/q_\lambda})} \right).
    \label{eqn:sondheimer_sum}
\end{equation}
We note that this approximate formula agrees remarkably well with direct numerical integration over $n$ in Eq.~\eqref{eq:Sigma_0_diffusive}. Let us analyze Eq.~(\ref{eqn:sondheimer_sum}) in more detail.
When $d \rightarrow \infty $, the second term can be neglected and we recover the three-dimensional result. For $\delta=0$ (the specular scattering limit), the second term also disappears since $\tilde{R}=1$, and we recover the Drude formula, Eq.~(\ref{eq:density_quasi2d}). For finite $\delta$, Eq.~(\ref{eqn:sondheimer_sum}) can be viewed as the quantum generalization of the semiclassical Sondheimer formula in Eq.~(\ref{eq:sondh_full}). It reduces to the semiclassical result upon replacement of summation over $\lambda$ with an integration and treating $\tilde{R}$ as a constant. 
This replacement is justified by assuming $k_F d \gg 1$ and $d/l \gg 1/(k_F d)$. The second condition is particularly important because the exponent varies on the scale of $d/l$, and discrete summation effects may still be significant (even when $k_F d \gg 1$) if the spacing in $q_\lambda$ exceeds this scale. Thus, the $s=0$ harmonic of the quantum formalism successfully reproduces all the results obtained from the Boltzman formalism, including Sondheimer oscillations, and allows us to make predictions in the ultrathin limit. 

\begin{figure}[h]
    \begin{minipage}[h]{0.45\linewidth}
    \center{\includegraphics[width=1\linewidth]{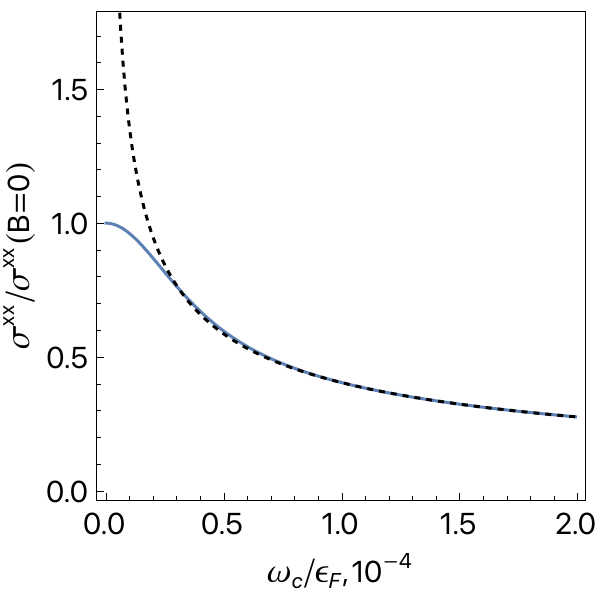}}
    \end{minipage} 
\caption{Longitudinal conductivity $\sigma^{xx}$ in the absence of bulk impurities ($1/\tau =0$), and at weak magnetic fields. The blue line depicts direct numerical integration in Eq.~\eqref{eq:conductivity_general}. The dashed black line shows the semiclassical limit obtained by replacing the sum in Eq.~\eqref{eq:Sigma_0_1/tau=0} with an integral. Parameters: $k_F d=100$, $\delta k_F=0.5$, $\tau^{-1}=0$.}
\label{fig:conductivity_nonosc}
\end{figure}

Let us now consider Eq.~\eqref{eqn:sondheimer_sum} in the limit  $1/\tau=0$ corresponding to the absence of bulk impurities
\begin{equation}
     \lim\limits_{1/\tau\rightarrow 0}\sigma^{xx}_0=\frac{\delta l_B^4}{\pi d^2 }\sum_{\lambda=1}^{\lambda_{\rm max}} \frac{q_\lambda^2\left(k_F^2-q_\lambda^2\right) (1+q_\lambda^2\delta^2)}{4q_\lambda^2 \delta^2+(1-q_\lambda^2\delta^2)^2\sin^2\left(\frac{d }{2q_\lambda l_B^2}\right)} \;\sin^2\left(\frac{d }{2q_\lambda l_B^2}\right)\;.\label{eq:Sigma_0_1/tau=0}
\end{equation}
This result is a quantum generalization of the classical expression in Eq.~\eqref{eq:Sigma_xx_1/tau=0_classical}. We emphasize that in deriving Eq.~\eqref{eq:Sigma_0_1/tau=0}, we neglected higher-order harmonics arising from the Poisson resummation over Landau levels in Eq.~\eqref{eq:conductivity_general_Q_0}; we will return to these contributions shortly. Since we also took the clean limit $\omega_c\tau\rightarrow \infty$, the validity of Eq.~\eqref{eq:Sigma_0_1/tau=0} should be restricted to the regime $T>\omega_c$, where Shubnikov–de Haas oscillations are suppressed by additional factors of $R_T$, see Eq.~\eqref{eq:R_T_s}. At the same time, the discrete $k_z$ levels remain resolved as long as $T/\epsilon_F \leq 1/d^2k_F^2$ (for small $\lambda$) or $T/\epsilon_F \leq 1/dk_F$ (for $q_\lambda \approx k_F$), whereas for higher temperatures the sum can be replaced by the integral, and we reproduce Eq.~\eqref{eq:Sigma_xx_1/tau=0_classical}. Furthermore, taking the zero field limit $B=0$ in the derived formula yields a finite answer, contrary to the semiclassical case. This is demonstrated in Fig.~\ref{fig:conductivity_nonosc}: the semiclassical conductivity given by Eq.~(\ref{eq:sondh_full_T}) diverges logarithmically, whereas the quantum conductivity stays finite. After expanding Eq.~\eqref{eq:Sigma_0_1/tau=0} up to the second order in $B$, we find 
\begin{equation}
 \lim\limits_{1/\tau \rightarrow 0}\sigma^{xx}_0=\sum_{\lambda=1}^{\lambda_{\rm max}} \frac{k_F^2-q_\lambda^2}{16 \pi \delta q_\lambda^2} \left( 1+q_\lambda^2 \delta^2\right) - B^2 \sum_{\lambda=1}^{\lambda_{\rm max}} \frac{d^2(k_F^2-q_\lambda^2)}{2^6 \pi\delta q_\lambda^4}\Big(1+q_\lambda^2\delta^2\Big)\left(\frac{1}{3}+ \frac{(1-q_\lambda^2\delta^2)^2}{4q_\lambda^2\delta^2}\right) +\mathcal{O}(B^4).\label{eq:mag_cond_B^2}
\end{equation}
The first term in this expansion, corresponding to the zero magnetic field case ($B=0$), was originally derived in Ref.~\cite{Sheng1995}. The second term corresponds to a negative quadratic magneto-conductivity at small fields, see Fig.~\ref{fig:conductivity_nonosc}.

Next, we consider the oscillatory terms that emerge from higher-order harmonics ($s \ge 1$) after applying the Poisson summation formula over Landau levels:

\begin{equation}
 \sigma^{xx}_{{\rm osc}}=\sum_{s=1}^\infty (-1)^s \frac{d}{(2\pi l_B^2)^2}\int dn n \Re \left( \frac{\overline{\Upsilon(n,z,z')\Upsilon^*(n-1,z,z')}}{\kappa(n) \kappa(n-1)(1-RT(2d,n))(1-R T(2d,n-1)^*)}\right)e^{2 \pi i s n}.
    \end{equation}

The average over $z,z'$ was already calculated in Eq.~(\ref{eq:average_Gamma}). After evaluating the residues at the poles given in Eq.~(\ref{eq:poles}) we obtain:
\begin{equation}  \sigma^{xx}_{{\rm osc}}= \Re \frac{\tau_B/m}{1+i\omega_c\tau_B}\sum_{\lambda=1}^{\lambda_{max}} \frac{k_F^2-q_\lambda^2}{2\pi d} \left[1-\frac{q_\lambda(1-\tilde{R})(1-e^{-\zeta k_F/q_\lambda})}{\zeta k_F(1-\tilde{R} e^{-\zeta k_F/q_\lambda})} \right]  \sum_{s=1}^{\infty}(-1)^s e^{-
\frac{\pi s}{\omega_c\tau_B}}\; \Re e^{\pi i s \frac{k_F^2-(q_\lambda+i\kappa_2)^2}{m\omega_c}} .
\label{eqn:sondheimer_sum_quant}
\end{equation}
Note, that we have reinstated the oscillatory scattering rate $1/\tau_B$ here. Moreover, we emphasize that $\kappa_2$ is important for Sondheimer oscillations and cannot be neglected since it encodes the damping effects of the surface. At $\delta=0$, Eq.~(\ref{eqn:sondheimer_sum_quant}) coincides with Eq.~(\ref{eqn:cond_zero_Q_arb_d}) derived in the previous section.

Using the Poisson summation formula and the saddle point approximation, we derive an expression for oscillations at finite boundary scattering length $\delta$, generalizing the result in the specular limit, see Eq.~(\ref{eq:cond_full_simplified}):
\begin{equation}
  \frac{\sigma^{xx}_{\rm osc}}{\sigma_0}= \frac{3\tau_B}{2   \tau}\sqrt{\frac{\omega_c}{2 \epsilon_F}}\sum_{s=1}^{\infty}  \frac{(-1)^s}{\sqrt{s}} e^{-
  \frac{\pi s}{\omega_c\tau_B}}\sum_{p=0}^{p_s} (2-\delta_{p,0} )
   A_{p,s}  
r_{p,s}^{p} \cos\left( \frac{2 \pi  s\epsilon_F}{\omega_c} +  \frac{ p^2 d^2}{\pi s l_B^2} -\frac{\pi}{4}\right),
\label{eq:cond_full_simplified_2}
\end{equation}
where $p_s=\pi k_F s/(B d)$ and the amplitude $A_{p,s}$ and suppression factor $r_{p,s}$ are given by
\begin{equation}
A_{p,s}=    \left(1-\frac{p^2 d^2}{\pi^2 s^2 k_F^2 l_B^4} \right) \left[ \frac{1}{1+ \omega_c^2 \tau_B^2}- \Re \frac{p \omega_c\tau_B(1-r_{p,s})(1-e^{-\frac{i\pi s}{p} -\frac{\pi s }{p\omega_c\tau_B}})}{\pi s(1+i\omega_c\tau_B)^2(1-r_{p,s} e^{-\frac{i\pi s}{p} -\frac{\pi s }{p\omega_c\tau_B}})}   \right] ,\quad r_{p,s}= \left(\frac{1- \delta k_F p/p_s}{1+\delta k_F p/p_s }\right)^2\;.
\label{eq:A_def_p_s}
\end{equation}
In addition to the Dingle factor $R_D$, which suppresses higher harmonics in $s$, the $p$-harmonics experience an additional suppression through the factor $r_{p,s}^p$. This term is absent in the specular limit ($\delta=0$) and reflects the inevitable loss of coherence caused by multiple scattering from the boundaries. Furthermore, the sum in $p$ is bounded by a hard cutoff $p<p_s$, which indirectly causes the Dingle-factor to suppress $p$-harmonics as well. Importantly, the frequencies of the oscillations do not depend on the boundary scattering length $\delta$, and can be understood from the semiclassical analysis, see Eq.~(\ref{eqn:self_energy_full}). Equations (\ref{eqn:sondheimer_sum}) and (\ref{eq:cond_full_simplified_2}) provide the most general results for the conductivity oscillations when the number of Landau levels is large. We analyze these oscillations in more detail in the next section, focusing on the first few harmonics.

\subsection{The limit of a few filled Landau levels}\label{subsec:fewLL}

So far, all the analysis has been performed in the limit of many Landau levels, where the Poisson summation formula is valid. In this subsection, we focus on the opposite case when only a few Landau levels are occupied. While the SCBA approximation is not exact in this limit, the deviations are milder than in the two-dimensional case. The expression for the self-energy, similar to Eq.~(\ref{eq:self_energy_poisson}), is:

\begin{equation}
   \overline{\Sigma}_v^R =\frac{\omega_cd}{2\pi  \lambda_{\rm max} \tau} \sum_{n=0}^{\infty}  \frac{m }{i \kappa(1- R e^{2 i \kappa d})}\left( 1+Re^{2i \kappa d}+i \sqrt{R}\frac{e^{2 i \kappa d}-1}{\kappa d}\right),
\end{equation}
 where $\kappa^2=k_F^2-2 m \omega_c(n+1/2)-2 m \overline{\Sigma}_v^R$ and the reflection coefficient $R=(1-\kappa \delta)^2/(1+\kappa \delta)^2$. 

In the limit of many modes $k_F d \gg1$ and strong diffusive scattering $R e^{2 i \kappa d}\ll1$ we obtain the following expression:

\begin{equation}
   \overline{\Sigma}_v^R =-\frac{i}{\tau}  \sum_n\frac{d}{2 \pi l_B^2 \lambda_{max} \kappa} \left(1+2 R^2 e^{2 i \kappa d}\right).
\end{equation}
We can further approximate $\kappa\approx \bar{\kappa}+i m /(2 \tau \bar{\kappa})$, where $\bar{\kappa}^2=k_F^2-2 m \omega_c(n+1/2)$ to highlight the oscillating part. The first term of the self-energy does not oscillate and describes Van-Hove singularities that occur each time $\bar{\kappa}=0$ and a new Landau level passes the Fermi energy. The second term oscillates with the frequency $2 \bar{\kappa} d$, which results from the quantization of levels in $k_z$ momentum. In contrast to Section \ref{sec:Sigma_finite_B}, where complex interferences between the discrete levels took place, the oscillations described above simply follow from the discreteness of individual levels. We further note that such oscillations are suppressed by the factor $R^2$ once the boundary scattering becomes diffusive. 

The most general expression for the conductivity is given by Eq.~(\ref{eq:conductivity_general}) with the Green's functions for the diffusive boundary derived in Eq.~(\ref{eq:full_G_with_boundary_finite_B}). The averages over $z$ and $z'$ can be performed analytically, however, the final expression appears to be too cumbersome to analyze. Instead, we will focus on the specular case with $\delta=0$, where the Kubo formula simplifies, see Eq.~(\ref{eq:conductivity_general_Q_0}). After applying the Poisson summation formula to evaluate the sum over $\lambda$, we obtain:
\begin{equation}
    \sigma^{xx}_v=\frac{3\omega_c^2}{4 \epsilon_F  } \frac{\sigma_0}{\omega_c \tau\sqrt{\omega_c^2\tau^2+1}}\sum_{n=0}^{\infty} (2n+1) \Re\left(\frac{1}{v_F \kappa} \right) \Re \left(\frac{1+e^{2 i \kappa d }}{1-e^{2 i \kappa d }}e^{+i \arctan (1/(\omega_c \tau))} \right)\;.
\label{eq:conductivity_fewn_Q_0}
\end{equation}

Similar to the self-energy, the non-oscillating part of the conductivity shows peaks each time a new Landau level crosses the Fermi energy. The frequency of the oscillations is $2  \bar{\kappa} d $ which coincides with the oscillations of self-energy and is caused by the quantization of $k_z$ momentum.

The Lifshitz-Kosevich factor acquires a nontrivial temperature dependence:
\begin{equation}
R_T(n,p)= \int d \omega \frac{\partial n_F(\omega)}{\partial \omega} e^{2  i\sqrt{k_F^2-2m \omega-2m\omega_c (n+1/2)} d p }\approx\frac{2 \pi p T m d/\bar{\kappa} }{\sinh \left(2 \pi p T m d /\bar{\kappa} \right)}\;.
\end{equation}
This result is obtained under the condition $T\ll \epsilon_F,\omega_c$ by expanding the square root. Since $ \Delta \epsilon=\pi \bar{\kappa}/(md)$ is the energy spacing between the discrete levels at the Fermi energy, the Lifshitz-Kosevich factor simply dampens the oscillations once the temperature exceeds this scale.
\section{Discussion}
\label{sec:disc}
In this section we summarize our findings and discuss the physical interpretation of the derived formulas. We begin with the case of many occupied Landau levels, where the Sondheimer contribution to the conductivity is given by Eq.~(\ref{eqn:sondheimer_sum}) while the additional quantum terms are described by Eq.~(\ref{eq:cond_full_simplified_2}). As a simple application of this general formula, we explicitly consider three lowest-order contributions describing the semiclassical Sondheimer oscillations ($s=0$), first harmonic of the Shubnikov–de Haas (SdH) oscillations ($s=1,p=0$), and the first harmonic associated with the discretization effects in the $z$-direction ($s=1,p=1$):
\begin{equation}
\begin{split}
 \frac{\sigma^{xx}_{\rm osc}}{\sigma_{v,0}^{xx}} &=\frac{3 (l/d)^3}{1+\omega_c^2 \tau^2}e^{-d/l} \bigg(1-\bigg(\frac{1-k_F \delta}{1+k_F \delta}\bigg)^2\bigg)^2\cos \left(\omega_c \tau\frac{ d}{l} +4 \arctan (\omega_c \tau)\right) \quad\quad  \quad\quad\quad\quad\text{"Sondheimer oscillations"}\\
 &-  \frac{    1+5\omega_c^2 \tau^2  }{2  (1+\omega_c^2 \tau^2)} \sqrt{\frac{\omega_c}{2\epsilon_F}}
R_D\cos \left(   \frac{2 \pi\epsilon_F}{\omega_c}  -\frac{\pi}{4}\right)\quad \quad\quad\quad\quad\quad\quad\quad\quad\quad\quad\quad\quad\quad\quad\quad\text{"SdH (extremal orbit)"}\\
&- \theta \left(\pi k_F l_B^2-d\right)\Big(\frac{1{+}5 \omega_c^2 \tau^2}{ 1{+}\omega_c^2 \tau^2}- \frac{3d^2}{\pi^2 k_F^2 l_B^4} \Big) 
\sqrt{\frac{\omega_c}{2\epsilon_F}} R_D \left(\frac{1-\delta B  d/\pi }{1+\delta B  d/\pi } \right)^{2}\cos \left(  \frac{2 \pi\epsilon_F}{\omega_c} +  \frac{d^2}{\pi l_B^2 } -\frac{\pi}{4}\right),\;\text{"SdH ($k_z$ quantization)"}\\
\end{split}
\label{eq:cond_full_full_sigma}
\end{equation}
where $\sigma_{v,0}^{xx}$ is the Drude conductivity in a magnetic field, given by Eq.~(\ref{eq:density_quasi2d}), and $R_D=e^{{-}\frac{\pi}{\omega_c\tau}}$ is the Dingle factor. Let us analyze Eq.~(\ref{eq:cond_full_full_sigma}) in more detail. The first term corresponds to the semiclassical Sondheimer oscillations and reproduces Eq.~(\ref{eqn:osc}) obtained from the Boltzmann formalism.
The amplitude of these oscillations decays exponentially with the thickness of the sample, therefore, it can be observed only when the thickness of the sample is comparable to the mean free path. It also decays as $1/B^4$ with the magnetic field, which was confirmed in early experiments \cite{Grenier1966}. Finally, the reflection coefficient $R$, which in the Boltzmann formalism was introduced empirically, now has a microscopic origin and depends on the boundary scattering length as $R=(1-k_F \delta)^2/(1+k_F \delta)^2$. In the case of fully specular scattering ($\delta=0$), one finds $R=1$, and the Sondheimer oscillations disappear. This is compatible with previous analyses, since Sondheimer oscillations are a semiclassical effect that arises from boundary roughness. The frequency of Sondheimer oscillations is $\omega_c d/v_F$ which scales linearly with the magnetic field $B$. 

The second term corresponds to the first harmonic of the Shubnikov-de-Haas oscillations. This contribution is fully quantum in origin and can not be derived from the Boltzmann formalism. The amplitude of the SdH oscillations is suppressed by the Dingle factor $R_D$ at weak magnetic fields ($\omega_c \tau \ll 1$), where the individual Landau levels could not be resolved. It is further suppressed by $\sqrt{\omega_c/\epsilon_F}$ factor typical for the 3D case. Interestingly, the oscillation amplitude is independent of the boundary scattering length $\delta$, since the dominant contributing orbits have $k_z=0$ and thus do not scatter from the boundaries. The frequency of SdH oscillations is $2\pi \epsilon_F/\omega_c$ and thus scales as $1/B$ with magnetic field, which makes these oscillations readily distinguishable from the semiclassical Sondheimer oscillations.

 \begin{figure}[h!]
    \begin{minipage}[h]{0.30\linewidth}
    \center{\includegraphics[width=1\linewidth]{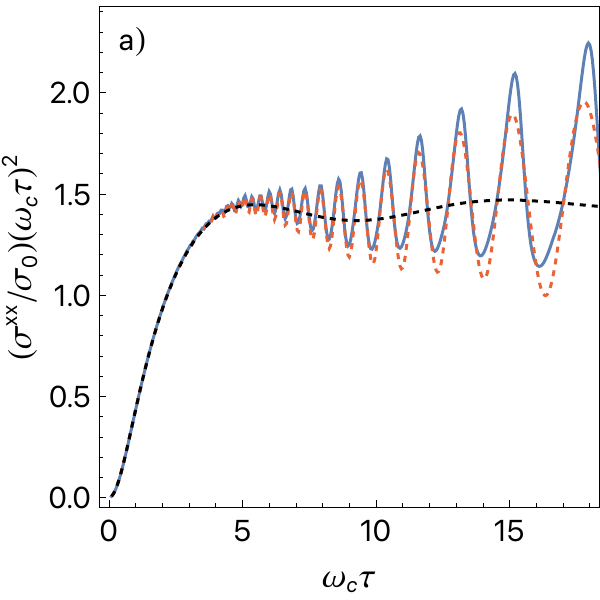}}
    \end{minipage} 
    \hfill
      \begin{minipage}[h]{0.32\linewidth}
    \center{\includegraphics[width=1\linewidth]{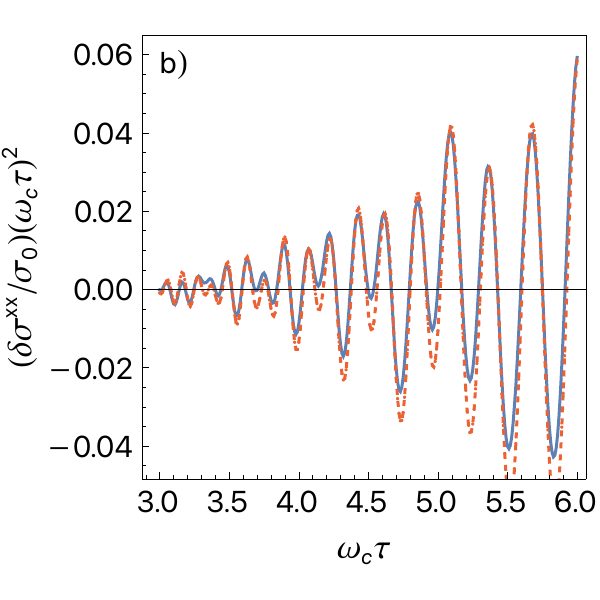}}
    \end{minipage} 
    \hfill
      \begin{minipage}[h]{0.32\linewidth}
    \center{\includegraphics[width=1\linewidth]{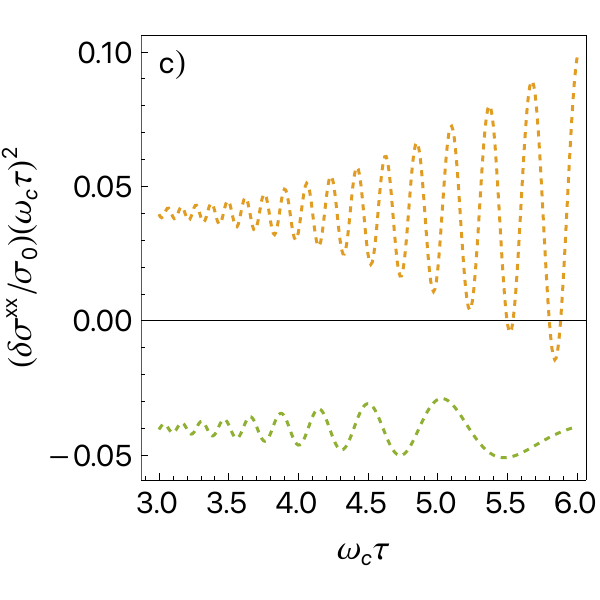}}
    \end{minipage}     
\caption{(a) Longitudinal conductivity $\sigma^{xx}$ computed by direct numerical integration in Eq.~\eqref{eq:conductivity_general} (blue line), Sondheimer part of the conductivity in Eq.~(\ref{eqn:sondheimer_sum}) (dashed black line),  analytic formula capturing both $s=1,p=0$ and $s=1,p=1$ harmonics (dashed red line). (b) Enlarged part of subfigure (a) focusing on size effects at intermediate field strength. The non-oscillatory contribution has been subtracted. (c) Two separate contributions to the oscillating part of $\sigma^{xx}$. The orange curve corresponds to the $s=1, p=0$ harmonic, whereas the green line shows $s=1, p=1$ harmonic. The parameters used are:  $T/\epsilon_F=1/100$, $\delta=0.5/k_F$, $d/l=0.5$, $k_F d=100$.
}
\label{fig:conduct_finite_T_3}
\end{figure}

The third term corresponds to the interference between Shubnikov-de-Haas oscillations and discretization effects in the $z$-direction. Similar to the pure SdH contribution, this term is fully quantum in origin and reflects the quantization of the $k_z$ momentum. Quasi-classically, it corresponds to finite $k_z$ trajectories that are reflected once from each surface before returning to their initial position, see Section \ref{sec:Sigma_finite_B} for more details. Because surface scattering is involved, the amplitude of the oscillations are additionally suppressed by the factor $\left((1-\delta B  d/\pi)/(1+\delta B  d/\pi ) \right)^{2}$ in addition to the Dingle factor $R_D$. Furthermore, the term effectively disappears once the sample thickness exceeds $ d> \pi k_F l_B^2$, since the discretizations effects become negligible. The right-hand side is proportional to the Larmor radius, and the inequality indicates that the electron accumulates a phase exceeding $2\pi$ after undergoing reflections from both surfaces. The oscillation frequency is also altered: apart from the conventional SdH $B$-dependence through $2\pi \epsilon_F/\omega_c$, there is an additional phase shift $d^2/(\pi l_B^2 ) $ linear in magnetic field $B$. This shift arises because the quantization effects displace the extremal cross sections from the equator, modifying the enclosed area. As shown in Section~\ref{sec:Sigma_finite_B}, applying the Bohr quantization condition reproduces the frequency shifts associated with $k_z$ quantization.

Fig.~\ref{fig:conduct_finite_T_3} compares the analytic approximation for conductivity oscillations in the presence of boundary scattering in Eq.~\eqref{eq:cond_full_full_sigma} with the general numerical expression in Eq.~\eqref{eq:conductivity_general}. In contrast to the $\delta=0$ regime depicted in Fig.~\ref{fig:conduct_1}, the smooth component (black, dashed line $s=0$) now exhibits slow modulations characteristic of Sondheimer oscillations. Superimposed on this are quantum oscillations resulting from in-plane and $k_z$ quantization. The latter emerge due to the finite film thickness when the condition $d/l\sim 1$ is met. The analytical formula works well for small $\omega_c \tau$ but deviates slightly from the full numerical expression at larger $\omega_c \tau$ when higher harmonics become relevant. Fig.~\ref{fig:conduct_finite_T_3} (c) shows two separate contributions from $s=1,p=0$(pure SdH oscillations) and $s=1, p=1$ (SdH + $k_z$ quantization) harmonics: they have different frequencies given by the last two terms of Eq.~\eqref{eq:cond_full_full_sigma}. Additionally, the $p=1$ harmonic is suppressed by the factor $r_{1,1}$ compared to the specular ($\delta=0$) case.

 \begin{figure}[h]
    \begin{minipage}[h]{0.45\linewidth}
    \center{\includegraphics[width=0.9\linewidth]{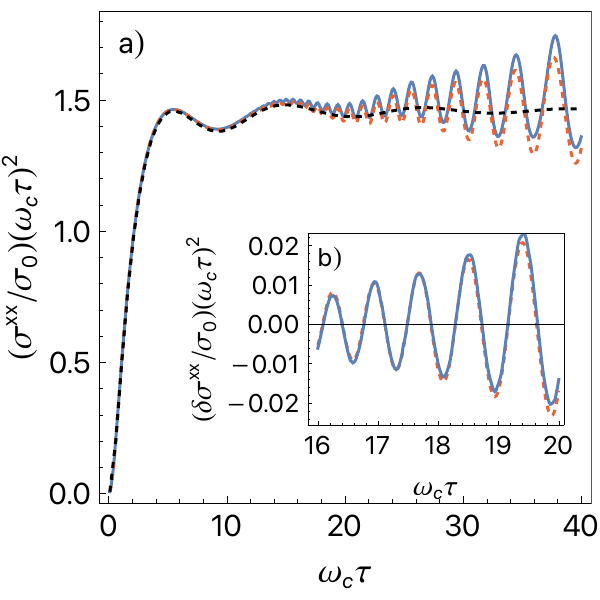}}
    \end{minipage} 
            \hfill
     \begin{minipage}[h]{0.45\linewidth}
    \center{\includegraphics[width=0.9\linewidth]{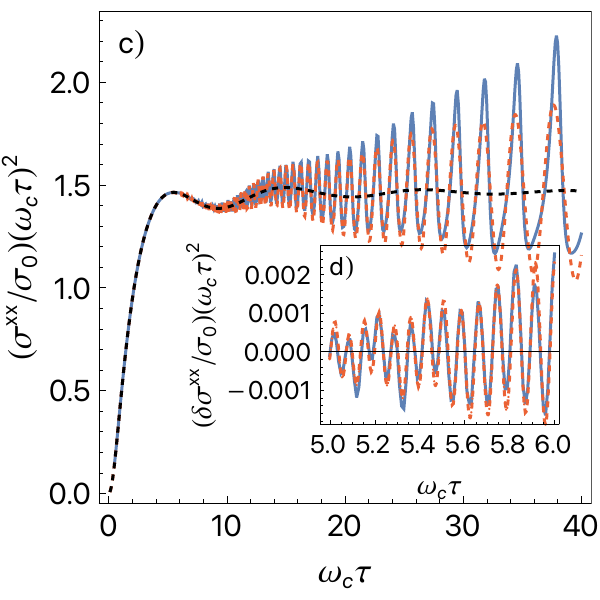}}
    \end{minipage} 
\caption{Longitudinal conductivity $\sigma^{xx}$ computed by direct numerical integration in Eq.~\eqref{eq:conductivity_general} (blue line). The black dashed line depicts the classical (Sondheimer) part of the conductivity, given by Eq.~(\ref{eqn:sondheimer_sum}). The red dashed line shows the result of the analytic formula \eqref{eq:cond_full_full_sigma} capturing both $s=1,p=0$ and $s=1,p=1$ harmonics. Insets (b), (d) show only oscillating parts of the conductivity. The parameters used for all figures are: $\delta k_F=0.5$, $d/l=0.5$, $k_F d=400$. The temperature $T/\epsilon_F=1/80$ was used for plots (a), (b), whereas  $T/\epsilon_F=1/200$ was used for plots (c), (d).
}
\label{fig:conduct_finite_T_4}
\end{figure}

Next, we discuss how conductivity is modified at finite temperatures. Sondheimer oscillations remain largely intact as long as $d \omega_c/v_F \ll \epsilon_F/T$, and decay exponentially at stronger magnetic fields. The quantum terms acquire an additional Lifshitz-Kosevich factor $R_T(s)$, Eq.~\eqref{eq:R_T_s}, inside the sum, and decay exponentially once $T\gg\omega_c$. This is demonstrated in Fig.~\ref{fig:conduct_finite_T_4}(a) and (c), where SdH oscillations almost disappear at small $\omega_c \tau$ while the Sondheimer component remains intact as the temperature is raised, since $\epsilon_F/T\gg 1$ in most materials. This behavior is also consistent with recent experimental observations in high-purity cadmium films \cite{guo2025} and other clean thin-film systems.
As noted in \cite{Delft2021}, stability of Sondheimer oscillations at large temperatures could be useful in exploring the hydrodynamic regime, where the electron-electron collisions dominate. We can also estimate the temperature at which $k_z$ quantization effects vanish. The sum over $p$ harmonics is restricted by $p_s= \pi k_F s/(Bd)$ and at $s>\omega_c/T$ the amplitude is damped by the Lifshitz-Kosevich factor. Therefore, at $T\gg \epsilon_F/(k_F d)$, all nontrivial harmonics with $p>0$ are effectively exponentially suppressed. The result is physically intuitive, since $\epsilon_F/(k_F d)$ corresponds to the energy spacing between discrete $k_z$ levels at the Fermi energy. We also note that the analytical formula in Eq.~\ref{eq:cond_full_full_sigma}, modified by Lifshitz-Kosevich factors at finite temperature, describes the exact numerical result quite well, and thus could be used for many practical applications. 


Apart from describing the intricate nature of quantum oscillations in the constraint geometry, our theory also resolves the unphysical divergence in the longitudinal conductivity predicted by semiclassical Boltzmann theory in the limit of infinite bulk mean free path. In this regime, the conductivity is rendered finite by boundary scattering alone. The key result is given in Eq.~\eqref{eq:Sigma_0_1/tau=0}, a quantized generalization of the classical expression in Eq.~\eqref{eq:Sigma_xx_1/tau=0_classical}. Its weak-field expansion leads to a negative quadratic magnetoconductivity, with a prefactor that depends sensitively on film thickness, as shown in Eq.~\eqref{eq:mag_cond_B^2} and illustrated in Fig.~\ref{fig:conductivity_nonosc}.



 \begin{figure}[h!]
      \begin{minipage}[h]{0.45\linewidth}
    \center{\includegraphics[width=1\linewidth]{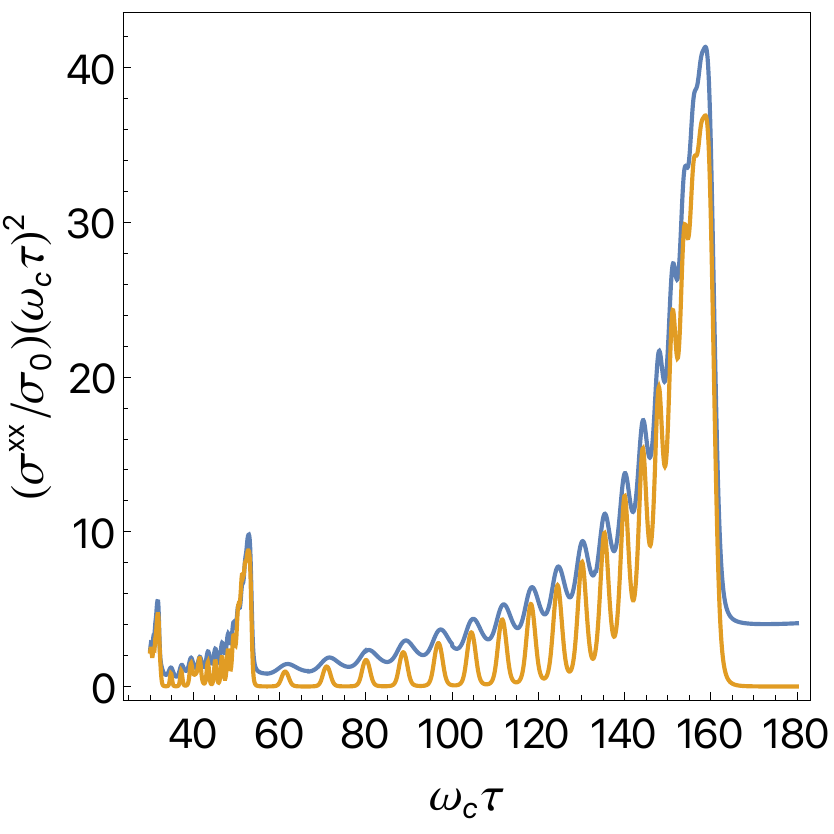}}
    \end{minipage}     
\caption{Longitudinal conductivity $\sigma^{xx}$ with few Landau levels present. At $\omega_c \tau \in [53.\overline{3},160]$ only one Landau level is present below the Fermi energy. The blue line corresponds to diffusive boundaries ($\delta k_F=0.5$) and the orange line corresponds to specular boundaries ($\delta=0$). Other parameters are chosen to resemble thin-film graphite, see \cite{Tianji2018}: $k_F d=80$, $d/l=0.5$,  $T/\epsilon_F=1/800$.  
}
\label{fig:experiment}
\end{figure}

In Sec.~\ref{subsec:fewLL}, we also considered the situation in which only a few Landau levels are filled. In the absence of disorder at the boundaries ($\delta=0$) the conductivity is described by Eq.~(\ref{eq:conductivity_fewn_Q_0}). Fig.~\ref{fig:experiment} shows the conductivity oscillations computed by numerically summing over Landau levels in Eq.~(\ref{eq:conductivity_general}).
There are several prominent features as conductivity varies with the magnetic field. Each time a new Landau level crosses the Fermi energy, i.e. when the condition $k_F^2/2m-\omega_c(n+1/2)=0$ is satisfied, the conductivity experiences an upturn. This results in conventional SdH oscillations with a characteristic $1/B$ periodicity. When only one Landau level is present, the conductivity oscillates with a single frequency $2d\sqrt{k_F^2-m \omega_c}$, coming from the discretization of levels in the $z$-direction. If $k_Fd\gg1$ the conductivity exhibits many oscillations before the magnetic field changes substantially. Therefore, the oscillations are linear in $\Delta B$ locally, with the corresponding frequency $d \Delta B/\sqrt{k_F^2-m \omega_c}$. As $\omega_c$ increases, the oscillation frequency also rises, as could be clearly observed in Fig.~\ref{fig:experiment}.
At smaller magnetic fields, second Landau level crosses the Fermi energy and adds an additional harmonic with the frequency $2d\sqrt{k_F^2-3m \omega_c}$. Thus, the resulting oscillations are a product of interference between the two frequencies, coming from different Landau levels. Finally, we note that the disordered boundaries further reduce the amplitude of the oscillations, as was rigorously demonstrated for the self-energy.

\section{Conclusions}
\label{sec:concl}

In this work, we have developed a quantum theory of magneto-transport in thin metallic films that extends the concept of Sondheimer oscillations into the fully quantum regime. By treating Landau quantization, surface scattering, and dimensional confinement along the magnetic field direction on an equal footing, our approach overcomes the limitations of semiclassical Boltzmann theory and enables a systematic exploration of the high-field, low-disorder limit where geometric effects in transport are significant. On the technical level, our calculations are based on the Kubo formula for conductivity, with both bulk and surface disorder treated within the self-consistent Born approximation (SCBA).

Our results reveal a rich interference pattern between Sondheimer oscillations, originating from diffusive boundary scattering in a constrained film geometry, and Shubnikov–de Haas (SdH) oscillations, which arise from Landau level quantization. The latter also exhibits additional harmonics due to momentum quantization along the magnetic field direction. These oscillatory phenomena coexist and interact in ways that are highly sensitive to the sample thickness, magnetic field strength, and boundary scattering conditions. Our most general expression for the oscillatory component of the conductivity is given in Eqs.~\eqref{eq:cond_full_simplified_2} and \eqref{eq:A_def_p_s}, where the result is structured as a double series over two types of harmonics: the $s$-harmonics correspond to in-plane Landau level quantization, while the $p$-harmonics arise from quantization of motion along the $z$-direction. The two lowest-order contributions in this expansion with small parameters $\omega_c \tau \ll 1$ and $l/d \ll 1$ are explicitly shown in Eq.~\eqref{eq:cond_full_full_sigma}, which serves as a minimal extension of Sondheimer’s original formula in Eq.~\eqref{eqn:osc}. Representative results are illustrated in Fig.~\ref{fig:conduct_finite_T_3} and the temperature dependence is shown in Fig.~\ref{fig:conduct_finite_T_4}. This formulation is particularly well-suited for the regime with many transverse modes, $k_F d \gg 1$. For systems with only a few quantized $k_z$-modes, equivalent expressions more appropriate to that limit are provided in Eqs.~\eqref{eqn:sondheimer_sum} and \eqref{eqn:sondheimer_sum_quant}.

To bring our work closer to experimental findings, we comment on two recent experiments conducted on thin-film graphite \cite{Taen2023} and in cadmium crystals \cite{guo2025}. The first experiment measured the longitudinal resistivity of clean thin graphite films under a strong magnetic field. The authors observed two distinct types of oscillations: the first appears at low magnetic fields and is periodic in $1/B$, while the second emerges in the range $B=10-25 \ T$ and is periodic in $B$. The frequency of the second type of oscillations was shown to be close to that predicted by Sondheimer. The experiment was carried out in the regime of a few filled Landau levels, where semiclassical theory is not sufficient and a full quantum description is required. To qualitatively interpret the experimental data, we apply the results of Sec.~\ref{subsec:fewLL} to compute the conductivity in a parameter regime close to the experimental conditions (see Fig. \ref{fig:experiment} to be compared with Fig. 1(d) of Ref. \cite{Taen2023}). In our classification, the first type of oscillations corresponds to a combination of conventional SdH oscillations and discretization effects, whereas the second type arises from the discreteness of $k_z$ levels when only one Landau level is present. We note that the frequency of the second type in our analysis is $d \Delta B/\sqrt{k_F^2-m \omega_c}$, defined with respect to a small magnetic field window $|\Delta B|\ll \epsilon_F-\omega_c/2$, and it increases with magnetic field. In future experiments, it would be interesting to observe this effect. Strictly speaking, our calculation is only illustrative in this case, as the Fermi surface of graphite is not spherical, and the SCBA eventually breaks down when only a few Landau levels are filled. Nevertheless, our analysis still captures the essential physics and could be easily extended to more general Fermi surfaces.

In the second paper \cite{guo2025}, the authors measure conductivity in thin cadmium crystals. In contrast to the previous case, the observed oscillations can be entirely attributed to classical Sondheimer oscillations, without any indication of the quantum SdH oscillations with frequencies proportional to $1/B$. One of the striking deviations from the conventional Sondheimer formula observed in \cite{guo2025} is a non-monotonic behavior of the amplitude at low magnetic fields. We believe that this misalignment between theory and experiment could be explained either by finite temperature effects (as discussed in Sec.~\ref{sec:sondh}) or by the peculiarities of the cadmium Fermi surface as well as anisotropic scattering rates. 

In the first scenario, finite temperature introduces an additional exponential decay to the amplitude: $B^{-3} e^{-B/B_0}$ where $B_0=2 m v_F \epsilon_F/(\pi  d T)$. In particular, $B_0\propto d^{-1}$, consistent with experimental observations. However, in this scenario the exponential suppression occurs at strong rather than weak magnetic fields, in contrast to the experiment.
The other scenario is explored in Appendix \ref{app:sondh}, where we generalize the semiclassical Sondheimer analysis to a non-quadratic Fermi surface with anisotropic relaxation times. We show that the experimentally observed non-monotonic behavior of the amplitude is reproduced by the modified semiclassical theory without invoking more subtle quantum effects.

Beyond the fundamental interest, our findings offer a practical framework for analyzing quantum transport in mesoscopic systems where both quantum coherence and geometric confinement are important. The theory provides new ways to extract information about surface scattering, mean free path, and confinement effects from experimental data. Our approach enables several natural extensions, such as exploring interference effects, studying bands with nontrivial topology, and analyzing Fermi liquid corrections, which we leave for future work.

\section*{Acknowledgements}
We acknowledge Eslam Khalaf, Subir Sachdev, and Igor Burmistrov for fruitful discussions. We are especially grateful to Leonid Levitov for bringing this problem to our attention and for discussions at the initial stages of this work. 

\appendix

\section{Spatial oscillations in the local density of states }
\label{app:spat_osc}

\begin{figure}[t!]
      \begin{minipage}[h]{0.45\linewidth}
    \center{\includegraphics[width=1\linewidth]{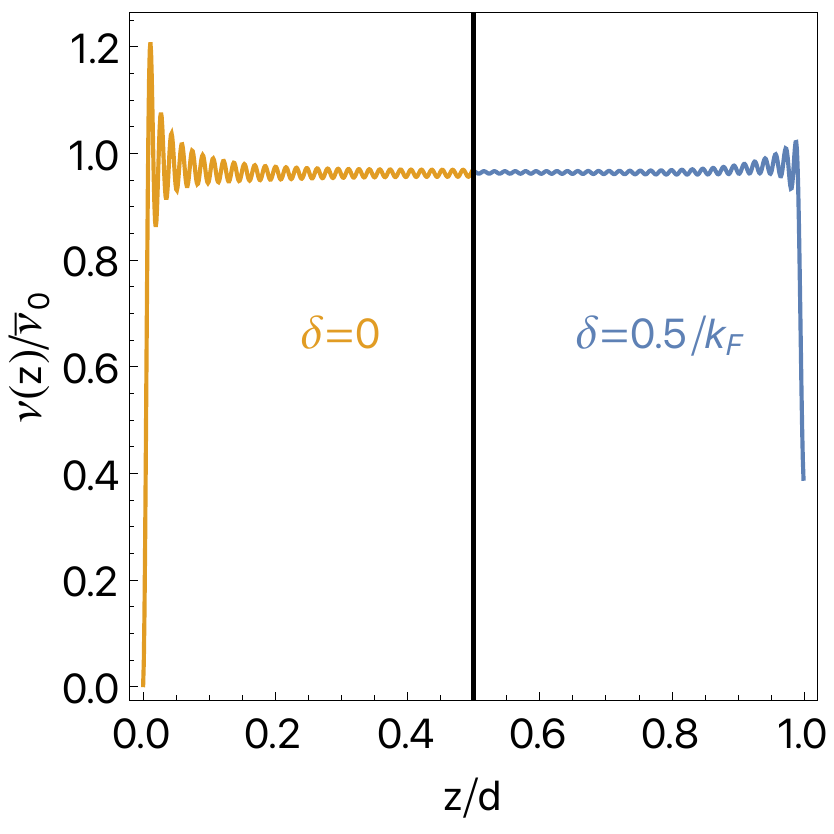}}
    \end{minipage}     
\caption{Friedel oscillations of the local density of states calculated with ($\delta\neq 0$) and without ($\delta= 0$) boundary disorder. Parameters: $k_F d=200$, $d/l=1/2$.  
}
\label{fig:ldos}
\end{figure}

The integral over $k_{\parallel}$ in Eq.~(\ref{eq:self_energy_B_0}) can be evaluated and we obtain

\begin{equation}
     \nu_0(z)=-\frac{1}{\pi}\int \frac{d^2k_{\parallel}}{(2\pi)^2}\Im G^R_{v}(k_\parallel,z,z)=\frac{ m}{\pi d}\sum_{\lambda=1}^{\lambda_{max}} \sin\left(q_\lambda z \right)^2=\frac{m}{4\pi d} \left(1+2 \lambda_{max}-\frac{\sin \left( \pi (1+2 \lambda_{max})z/d\right)}{\sin \pi z/d} \right),
\end{equation}
where $\lambda_{max}=[k_F d/\pi]$ denotes the number of modes in the $z$-direction. In the limit $k_F d \gg 1$ the formula simplifies to
\begin{equation}
    \nu_0(z)=\frac{ m k_F}{2\pi^2} \left(1-\frac{\pi}{2 k_F d} \frac{\sin 2 k_F z}{\sin \pi z/d} \right).
    \label{eq:ldos}
\end{equation}
The first term is independent of  $z$ and corresponds to the density of states in the three-dimensional system ($\bar{\nu}_0$). The second term oscillates with wavevector $2k_F$, and is attributed to Friedel oscillations arising from the presence of a boundary. Fig.~\ref{fig:ldos} illustrates the oscillatory behavior of the local density of states along the $z$-direction. The right panel in Fig.~\ref{fig:ldos} also shows that the Friedel oscillations are strongly suppressed by surface disorder.

\section{Comparison between first and self-consistent Born approximations}\label{sec:appendix_few_LLs}
In the main text, Eq.~({\ref{eq:self_en_full}) for the self-energy was obtained using the first Born approximation (i.e. just its first iteration was used). In the two-dimensional case, this approximation breaks down at $\omega_c \tau \gg1$ since Landau levels become well separated. In such cases, solving the full non-linear equation for $\overline{\Sigma}_v^R$ is required. However, in the three-dimensional case, the Landau levels disperse with $k_z$ momentum, which causes them to overlap even in the presence of weak disorder. This modification improves the validity of the first Born approximation. Fig.~\ref{fig:Born_approximation_comparison} compares the self-energy computed using the first Born approximation (blue line), with the fully self-consistent Born approximation, obtained by iteratively solving Eq.~({\ref{eq:self_en_full}) (green line). The difference between the two approaches is minimal. As expected, at small magnetic fields, only a few harmonics contribute, while at larger magnetic fields the oscillations become more involved.

 \begin{figure}[h]
    \begin{minipage}[h]{0.45\linewidth}
    \center{\includegraphics[width=1\linewidth]{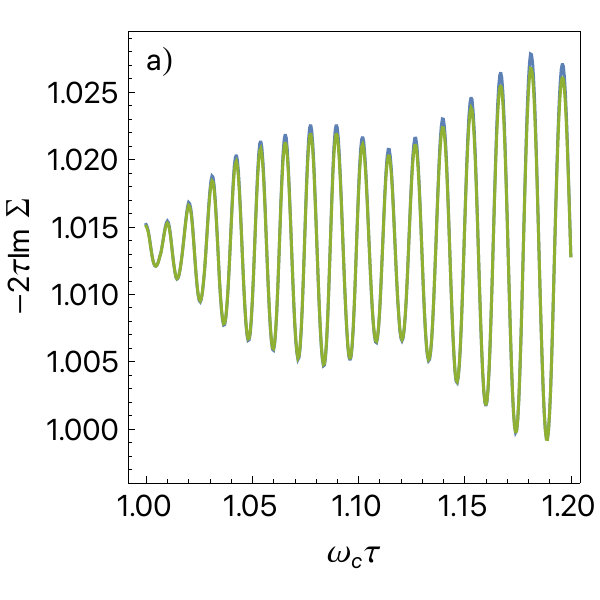}}
    \\a)
    \end{minipage} 
    \hfill
      \begin{minipage}[h]{0.45\linewidth}
    \center{\includegraphics[width=1\linewidth]{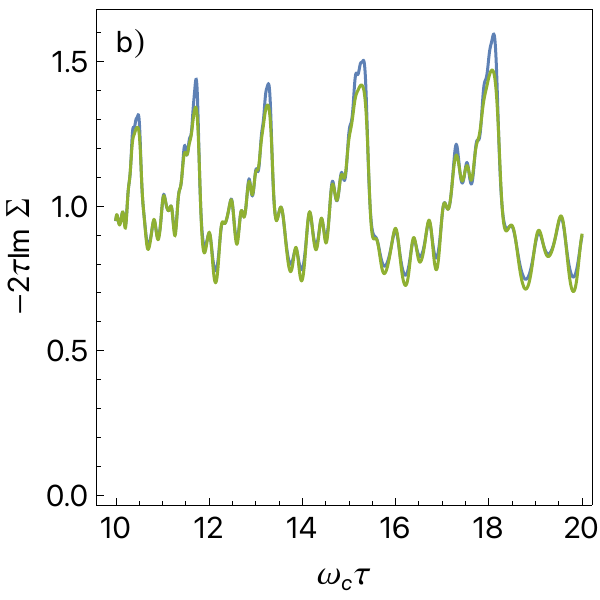}}
    \\b)
    \end{minipage}   
\caption{Self-energy obtained using the first Born approximation (blue line) and the full self-consistent Born approximation (green line). Parameters: $T/\epsilon_F=1/100$, $\delta k_F=1/2$, $d/l=1/2$, $k_F d=100$.
}
\label{fig:Born_approximation_comparison}
\end{figure}

\section{Generalization of the semiclassical Sondheimer theory }
\label{app:sondh}
In this section, we generalize semiclassical Sondheimer theory to the warped cylindrical Fermi surface with dispersion $\epsilon_k=(k_x^2+k_y^2)/(2m)+h(k_z)-\mu$ and anisotropic scattering rates $\tau_{\parallel}$ and $\tau_{\perp}$. After linearizing the distribution function around its equilibrium value
$f(\vec{v},z)=f_0(\vec{k})+(c_x(k_\perp,k_z,z) k_x+c_y(k_\perp,k_z,z) k_y)\partial f_0/\partial \epsilon$ Eq.~(\ref{eq:sondh_cx_cy}) is generalized as follows:

\begin{equation}
    \begin{dcases}
        &\frac{E_x}{m }-\frac{B}{m}c_y+v_z \frac{\partial c_x}{\partial z}=-\frac{c_x}{\tau_{\parallel}}-\frac{c_y}{\tau_{\perp}}\\
          &\frac{E_y}{m }+\frac{B}{m}c_x+v_z \frac{\partial c_y}{\partial z} =-\frac{c_y}{\tau_{\parallel}}+\frac{c_x}{\tau_{\perp}},\\
    \end{dcases}    
\end{equation}
where $v_z=d h(k_z)/d k_z$. The solution is
\begin{equation}
    c(\vec{k})=-\frac{E \tau_{\parallel}}{m}\frac{1}{1+i B \tau_{\parallel}/m- i \tau_{\parallel}/\tau_\perp} \left( 1+A(k_\perp,k_z) e^{-\frac{z}{\tau_{\parallel} v_z}(1+i B \tau_{\parallel}/m-i \tau_{\parallel}/\tau_\perp)}\right).
\end{equation}
We define $\beta=\tau_\parallel/\tau_\perp$ and compute the oscillatory correction to the longitudinal conductivity in the manner described in Section \ref{sec:sondh}. For simplicity, we assume that the scattering is fully diffusive ($R=0$) and that $h(k_z)=h(-k_z)$. The full expression for the conductivity is
\begin{equation}
 \sigma^{xx}=\Re\left( \frac{ \tau_\parallel}{2 \pi^2(1+i \omega_c \tau-i \beta)} \int_0^{k_z^{max}}  dk_z \left(1- \frac{   h'(k_z) \tau_\parallel}{d(1+i \omega_c \tau-i \beta)}  (1-e^{-d/(\tau_\parallel h'(k_z))(1+i\omega_c \tau-i \beta)})\right) (\epsilon_F-h(k_z)) \right).
\end{equation}
The oscillating part of the integral is given by
\begin{equation}
   \delta \sigma^{xx}=\Re\left( \frac{ \tau_\parallel}{2 \pi^2d(1+i \omega_c \tau_\parallel-i \beta)^2} \int_0^{k_z^{max}}  dk_z   e^{-d/(\tau_\parallel h'(k_z))(1+i\omega_c \tau_\parallel-i \beta)}(\epsilon_F-h(k_z))    h'(k_z) \tau_\parallel\right).
\end{equation}

To evaluate this integral at strong magnetic fields we use the stationary point method. Two cases should be considered separately: if $h''(k_z^i)=0$ has a solution, then the saddle point approximation yields
\begin{equation}
    \delta \sigma^{xx}=\Re\left( \frac{ \tau_\parallel^2(\epsilon_F-h)    h' }{2 \pi^2d(1+i \omega_c \tau_\parallel-i \beta)^2}   e^{-d/(\tau_\parallel h'))(1+i\omega_c \tau_\parallel-i\beta)}\sqrt{\frac{\pi \tau_\parallel}{2 d (1+i \omega_c \tau_\parallel-i \beta)}} \left( -\frac{h'''}{(h')^2}+\frac{2 (h'')^2}{(h')^3}\right)^{-\frac{1}{2}} \right) \Bigg|_{k_z=k_z^i},
\end{equation}
and the conductivity decays as $B^{-2.5}$ at strong magnetic fields. This exponent was first obtained by Gurevich \cite{gurevich1959}. If there is no saddle point, then one of the boundaries gives the leading contribution:
\begin{equation}
 \delta \sigma^{xx}= -\frac{ \tau_\parallel^4}{ 2 \pi^2 d^3(1+i \omega_c \tau- i \beta)^4}   e^{-d/(\tau h')(1+i\omega_c \tau_\parallel-i \beta)}  \frac{d}{d k_z} \Big[ \frac{(\epsilon_F-h)(h')^3}{h''}\Big] \frac{(h')^2}{h''} \Big|_{k_z=k_z^{max}}
\end{equation}
which gives the scaling $B^{-4}$ and coincides with Eq.~(\ref{eqn:osc}) in the case of a spherical Fermi surface $h(k_z)=k_z^2/(2m)$.

 \begin{figure}[t!]
      \begin{minipage}[h]{0.45\linewidth}    \center{\includegraphics[width=1\linewidth]{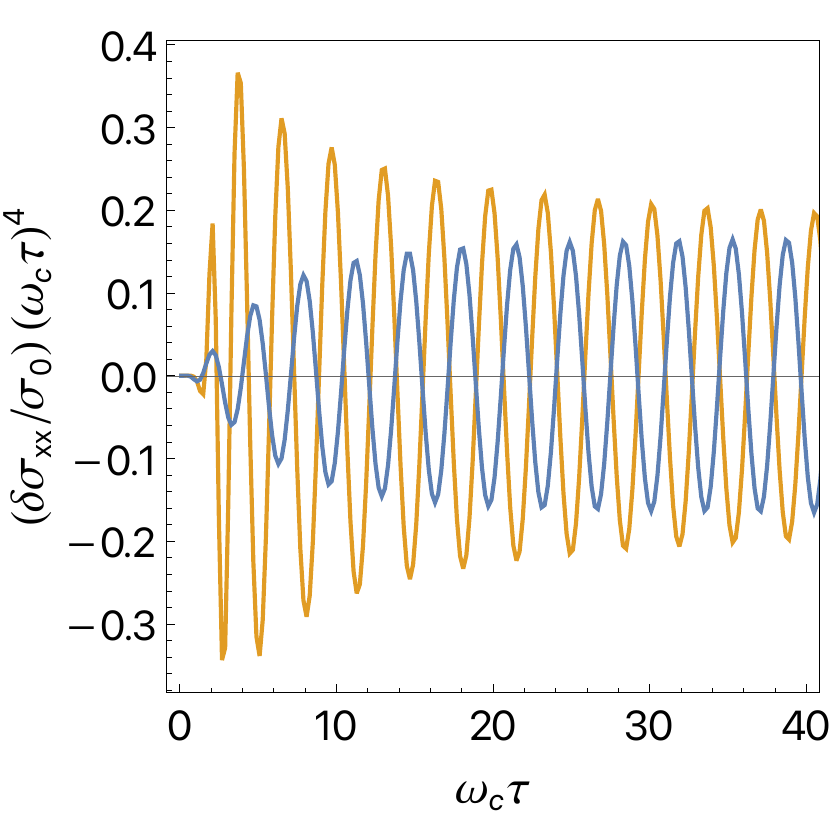}}
    \end{minipage}     
\caption{Oscillating part of longitudinal conductivity $\sigma^{xx}$ with the Fermi surface with $h(k_z)=k_z^2/((1+\alpha |k_z|))$. The blue line corresponds to $\beta=0.2$, and the orange line corresponds to $\beta=2$ (see the text for definitions). Parameters: $d/l=1$, $k_F d=9\times 10^4$, $k_F\alpha =1/2$.
}
\label{fig:experiment_kamran}
\end{figure}

The cadmium Fermi surface is well studied~\cite{Subedi2024} and a good approximation is $h(k_z)=k_z^2/(1+\alpha |k_z|)$, which makes the Fermi surface almost 'lens-like' near $k_z=k_F$. Fig.~\ref{fig:experiment_kamran} shows that the phase of the oscillations varies with magnetic field. Most notably, there is an upturn in the amplitude at low magnetic fields, similar to the experimental observations reported in Ref.~\cite{guo2025}.

\bibliography{bibliography}
\end{document}